\documentclass[prb,twocolumn,amsfonts,amssymb,amsmath,aps,superscriptaddress]{revtex4}
      
      \usepackage{graphicx} \usepackage{bm}

      \def\nn{\nonumber\\}
      
      \def\bk{\mathbf{k}} \def\ba{\mathbf{a}} \def\br{\mathbf{r}}
       \def\sB{{\scriptstyle{\mathbf{B}}}}
      \def\ssB{{\scriptscriptstyle{\mathbf{B}}}}
      
      \def\sK{{\scriptstyle{\mathbf{K}}}}
      \def\ssK{{\scriptscriptstyle{\mathbf{K}}}}
      \def\sQ{{\scriptstyle{\mathbf{Q}}}}
      \def\ssQ{{\scriptscriptstyle{\mathbf{Q}}}}
      \def\sR{{\scriptstyle{\mathbf{R}}}}
      \def\ssR{{\scriptscriptstyle{\mathbf{R}}}}

      \def\s{\sigma} \def\Jh{J_{\mathrm{H}}}
      
      \def\up{\uparrow} \def\down{\downarrow} \def\mS{\mathcal{S}}
      \def\mP{\mathcal{P}} \def\mD{\mathcal{D}} \def\mF{\mathcal{F}}
      \def\rK{\kappa} 

\begin{document}
 
 \title{Effective Interaction between the inter-penetrating Kagom\'e
   lattices in Na$_x$CoO$_2$}

 \author{Martin Indergand} \affiliation{Theoretische Physik,
   ETH-H\"onggerberg, CH-8093 Z\"urich, Switzerland} \author{Yasufumi
   Yamashita} \affiliation{Institute for Molecular Science, National
   Institutes of Natural Sciences, Okazaki 444-8585, Japan}
 \author{Hiroaki Kusunose} \affiliation{Physics Department, Tohoku
   University, Sendai 980-8578 Japan} \author{Manfred Sigrist}
 \affiliation{Theoretische Physik, ETH-H\"onggerberg, CH-8093
   Z\"urich, Switzerland}
 
 \date{\today}

 \begin{abstract}
   A multi-orbital model for a CoO$_2$-layer in Na$_x$CoO$_2$ is
   derived.  In this model the kinetic energy for the degenerate
   $t_{2g}$-orbitals is given by indirect hopping over oxygen, leading
   naturally to the concept of four inter-penetrating Kagom\'e
   lattices. Local Coulomb interaction couples the four lattices and
   an effective Hamiltonian for the interaction in the top band can be
   written in terms of fermionic operators with four different
   flavors. Focusing on charge and spin density instabilities, a big
   variety of possible metallic states with spontaneously broken
   symmetry are found. These states lead to different charge, orbital,
   spin and angular momentum ordering patterns.  The strong
   superstructure formation at $x=0.5$ is also discussed 
   within this model.
\end{abstract}


\maketitle
 \section{Introduction}\label{introduction}
 
 The layered Na$_x$CoO$_2$ has been initially studied for its
 extraordinary thermoelectric properties and for its interesting
 dimensional crossover.  \cite{tanakaT,terasaki,wang1,valla} But
 recently wider attention has been triggered by the discovery of
 superconductivity in hydrated Na$_{0.35}$CoO$_2$ and the discovery of
 an insulating phase in
 Na$_{0.5}$CoO$_2$.\cite{takada,schaak,foo,huang} Since then, various
 types of charge ordering phenomena in Na$_x$CoO$_2$ have been
 reported,\cite{zandbergen,shi,chen,huang2,wangNL,ray,gavilano,carretta,markus,bernhard,lupi,balicas,chen2,lupi2}
 but also strong spin-fluctuations and spin density wave transitions
 have been observed.
 \cite{boothroyd,helme,ihara,motohashi,sales,sugiyama1,sugiyama2,chou,caimi,luo,uemura,mendels}
  
 The material consists of CoO$_2$-layers where Co-ions are enclosed in
 edge-sharing O-octahedra. These layers alternate with the Na-ion
 layers with Na entering as Na$^{1+}$ and donating one electron each
 to the CoO$_2$-layer.
 
 The electronic properties are dominated by the $3d$-$t_{2g}$
 electrons of the Co-ions which form a two-dimensional triangular
 lattice. However, the spatial arrangement of the Na$^{1+}$-ions plays
 a crucial role too for the physics of this material.  There are two
 basic positions for the Na-ions, one directly above or below a
 Co-site and another in a center position of a triangle spanned by the
 Co-lattice. The metallic properties are unusual and vary with the
 Na-concentration and arrangement.
 
 A brief overview of the present knowledge of the phase diagram of
 Na$_x$CoO$_2$ leads to following still rough picture. The most
 salient and robust feature, at first sight is the charge ordered
 phase for $x=0.5$ separating the Na-poor from the Na-rich system. The
 Na-ions arrange in a certain pattern inducing an insulating magnetic
 phase in the CoO$_2$-layers below 50K.\cite{uemura} On the Na-poor
 side ($ x < 0.5 $) the compound behaves like a paramagnetic metal.
 When it is intercalated with H$_2$O superconductivity appears between
 $ x \approx 0.25 $ and $ x \approx 0.35 $. In several respects more
 interesting is the Na-rich side where one finds a so-called
 Curie-Weiss metal. Here the magnetic susceptibility displays a
 pronounced Curie-Weiss-like behavior after subtracting an underlying
 temperature independent part: $ \chi = C/(T-\Theta) $ where $ \Theta
 $ ranges roughly between -50 and -200 K depending on $x$, and the
 Curie constant is consistent with a magnetic moment in the range of 1
 - 1.7 $\mu_B $.  Note that deviations from the Curie-Weiss behavior
 have been observed at low temperatures.\cite{miyoshi} On the Na-rich
 side a transition at high temperature $ \sim $ 250 - 340 K has been
 observed and interpreted as crystal structure or charge ordering
 \cite{gavilano,sales,bernhard}.  For Na$_{0.75}$CoO$_2$ a magnetic
 transition occurs at 22 K and is most likely a commensurate spin
 density wave or ferrimagnetic order which is rather soft towards
 magnetic polarization. \cite{motohashi,sales,sugiyama1,sugiyama2}
 Interestingly this magnetic phase is metallic and has even a higher
 mobility than the non-magnetic phase.  For Na-content $ x\geq 0.75$
 several magnetic transition at a similar critical temperatures have
 been observed, but $\mu$SR-data suggest rather an incommensurate spin
 density wave order \cite{sugiyama1,sugiyama2,mendels}.
 
 The arrangement of the Na ions between the layers depends on the Na
 doping $x$ and several superstructures have been found,
 \cite{zandbergen,shi} the clearest evidence for the superstructure
 formation is at $x=0.5$ where the Na-ordering leads to a
 metal-insulator transition at low temperatures.
 \cite{foo,huang,huang2} But also away from $x=0.5$, NMR experiments
 indicate the existence of non equivalent cobalt sites and phase
 separation. \cite{ray,carretta}
 
 The complex interplay between Na-arrangement and the electronic
 properties poses a interesting problem. Various theoretical studies
 have mainly focused on single band models on the frustrated
 triangular lattice, in particular in connection with the
 superconducting phase ignoring Na-potentials.
 \cite{baskaran1,kumar,tanaka, ogata,wang2,carsten,ferraz} There is
 also work done on multi-orbital models \cite{maekawa,yanase} and
 density functional calculations have been
 performed.\cite{singh1,singh2,singh3,kunes,li,zhang,zhang2,johannes}
 According to LDA calculations, the Fermi surface lies near the top of
 the $3d$-$t_{2g}$-bands. They form a large hole-like Fermi surface of
 predominantly $a_{1g}$ character in agreement with ARPES experiments.
 \cite{hasan,yang,yang2} In addition the LDA calculations suggest,
 that smaller hole pockets with mixed $a_{1g}$ and $e'_g$ character
 exist on the $\Gamma-K$ direction on the Na-poor side.
 
 At the $\Gamma$ point the states with $a_{1g}$ and $e'_g$ symmetry
 are clearly split, but on average over the entire Brillouin zone the
 mixing between $a_{1g}$ and $e'_g$ is substantial.  Koshibae and
 Maekawa argued that the splitting at the $\Gamma$ point originates
 from the cobalt-oxygen hybridization rather than from a crystal field
 effect due to the distortion of the oxygen octahedra, because the
 crystal field effect in a simple ionic picture would lead to the
 opposite splitting of the $a_{1g}$ and $e'_g$ states.  \cite{maekawa}
 There is also spectral evidence, that the low-energy excitations of
 Na$_x$CoO$_2$ have significant O 2$p$ character.\cite{wu} Reproducing
 the LDA Fermi surface with a tight binding fit for the Co $t_{2g}$
 orbitals, it turns out that the direct overlap integral between the
 cobalt orbitals is much smaller than the indirect hopping integral
 over the oxygen 2$p$ orbitals.\cite{yanase} Therefore, it is
 reasonable to start with a three band tight binding model of
 degenerate $t_{2g}$ orbitals, where the only hopping processes are
 indirect hopping processes over intermediate oxygen orbitals.  This
 approximation provides an interesting system of four independent and
 inter-penetrating Kagom\'e lattices as it was already pointed out by
 Koshibae and Maekawa.
 
 Our study will be based on this model band structure which has a high
 symmetry. Within this model we examine various forms of order that
 could be possible from onsite Coulomb interaction.  The paper is
 organized as follows: In section \ref{tight-binding}, the
 tight-binding model and the concepts of Kagom\'e operators and pocket
 operators are introduced.  In section \ref{coulomb} an effective
 Hamiltonian for the local Coulomb interaction is derived and in
 section \ref{SU4order} this effective interaction is written in a
 diagonal form, by choosing an appropriate basis of SU(4) generators.
 In section \ref{breaksym}, the effects of small deviations from our
 simplified tight-binding model are discussed.  In section \ref{OP},
 all possible charge and spin ordering patterns of our model and the
 corresponding phase transitions, are shortly described.  In section
 \ref{possinst} the relevance of the above described collective
 degrees of freedom to Na$_x$CoO$_2$ is discussed by comparing the
 different coupling constants and by taking into account symmetry
 lowering effects.  In section \ref{superstructure} we apply our model
 to the Na-ordering observed at $x=0.5$ and we summarize and conclude
 in section \ref{discussion}.

 \section{Tight-Binding Model\label{tight-binding}}
 \begin{figure}[htbp]
   \includegraphics[width=\linewidth]{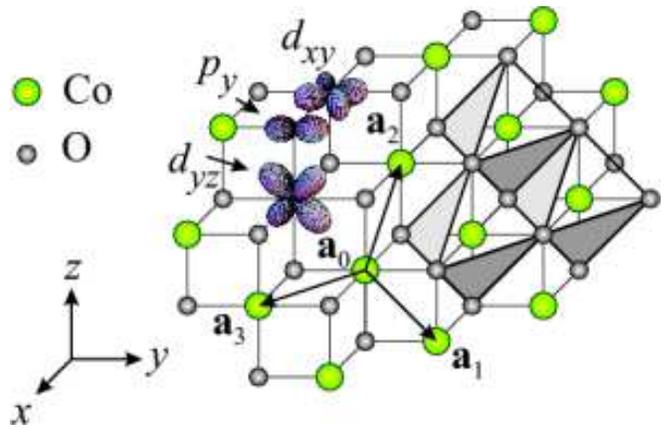}
   \caption{Schematic figure of a CoO$_2$ plane drawn with cubic unit cells. The edge-sharing of the oxygen octahedra around the Co ions is visualized. The edges of the cubes are oriented along the coordinate system $(x,y,z)$. The triangular lattice of the cobalt is spanned by the vectors $\ba_1, \ba_2, \ba_3$. ($\ba_1+\ba_2=-\ba_3$). $a=|\ba_i|$ is the lattice spacing. An oxygen $2p$ orbital and the cobalt $t_{2g}$ orbitals hybridizing with it by $\pi$-hybridization, are shown.}
   \label{CoO}
 \end{figure}
\begin{figure}[htbp]
  \includegraphics[width=\linewidth]{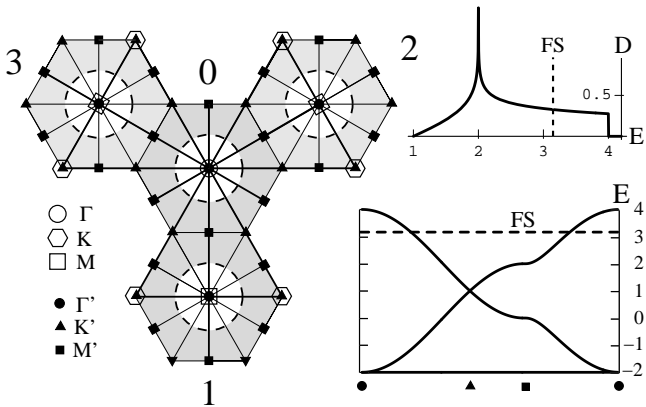}
   \caption{The original Brillouin zone (BZ) of the triangular lattice consists of four reduced BZs around  of the $\Gamma$ point (0) and the three M points (1,2,3). The  symmetry points of the reduced BZs, M', K' and  $\Gamma'$ are symmetry points for the tight binding model in Eq.~(\ref{tbham}) due to the higher periodicity of the bands. It is therefore sufficient to draw the bands along the lines $\Gamma$'-K'-M'-$\Gamma$'. The Fermi surface (FS) for $x=0.5$ lies at $E^1_{\mathbf{k}}\approx 3.16t$. The density of states per spin and per reduced Brillouin zone $D$ is given in units of $1/t$. It has a logarithmic singularity at $E=2t$.}
   \label{zones}
 \end{figure}

 We base our model on the assumption that the $3d$-$t_{2g} $-orbitals
 on the Co-ions are degenerate. Their electrons disperse only via $
 \pi $-hybridization with the intermediate oxygens occupying the
 surrounding octahedra (Fig.\ref{CoO}). As noticed by Koshibae and
 Maekawa the resulting electronic structure corresponds to a system of
 four decoupled equivalent electron systems of electrons hopping on a
 Kagom\'e lattice \cite{maekawa}.  The different sites, however, are
 represented by different orbitals. Each of the three orbitals $ \{
 d_{yz} , d_{zx} , d_{xy} \}$ on a given site participate in one
 Kagom\'e lattice, and the fourth Kagom\'e lattice has a void on this
 site.  The corresponding tight-binding model has the following form,
  \begin{equation}
   \label{tbham}
   H_{\mathrm{tb}}=\sum_{\bk\s}\sum_{mm'}\epsilon_{\bk}^{mm'}c^{\dag}_{\bk m\s}c_{\bk m'\s},
 \end{equation}
 where $c^{\dag}_{\bk m
   \s}=\frac{1}{\sqrt{N}}\sum_{\br}e^{i\bk\cdot\br}c^{\dag}_{\br m}$
 are the operators in momentum space of $c^{\dag}_{\br m\s}$ which
 creates a $t_{2g}$-orbital $(d_{yz}, d_{zx}, d_{xy})$ with index
 $m\in\{1, 2, 3 \}$ and spin $\s$ $\in\{\uparrow,\downarrow$\} on the
 cobalt-site $\br$. $N$ is the number of Co-sites in the lattice.
 \begin{equation}
   \label{tbmatrix}
    \hat{\epsilon}_{\bk}=\left(\begin{array}{ccc}-\mu&2t\cos(k_3)&2t\cos(k_2)\\
 2t\cos(k_3)&-\mu&2t\cos(k_1)\\
 2t\cos(k_2)&2t\cos(k_1)&-\mu\end{array}\right),
 \end{equation}
 with $k_i=\bk\cdot\ba_i$, cf.\ Fig.~\ref{CoO}. The hopping parameter
 $t=t_{pd}^2/\Delta>0$, where $t_{pd}$ is the hopping integral between
 the $p_y$ and the $d_{xy}$ or $d_{yz}$ orbital shown in
 Fig.~\ref{CoO}. $\Delta$ is the energy difference between the oxygen
 $p$ and the Co-$t_{2g}$ levels.  The diagonalization of the matrix
 $\hat{\epsilon}_{\bk}$ by a rotation matrix $\hat{O}_{\bk}\in SO(3)$
 \begin{equation}\label{rotmat}
\sum_{mm'}O^{im}_{\bk}\epsilon^{mm'}_{\bk}O^{jm'}_{\bk}=\delta_{ij}E^i_{\bk}
\end{equation}
results in the three energy bands
 \begin{eqnarray}
   \label{bands}
   E^{1}_{\bk}&=&t+ t\sqrt{1+ 8\cos(k_1)\cos (k_2)\cos (k_3)}-\mu\nonumber\\
   E^{2}_{\bk}&=&t- t\sqrt{1+ 8\cos(k_1)\cos (k_2)\cos (k_3)}-\mu\\
   E^3_{\bk}&=&-2t-\mu.\nonumber
 \end{eqnarray}
 These bands have the periodicity $E^l_{\bk+\ssB_j}=E^l_{\bk}$ , where
 the vectors $\sB_j$ are defined by
 \begin{equation}\label{scalerel}
 \ba_i\cdot\sB_j=\frac{2\pi}{\sqrt{3}}\sin(\theta_i-\theta_j)\quad i,j\in\{1\dots 3\}
 \end{equation}
 with $\theta_j=2\pi j/3$.  These three vectors $\sB_j$ connect the
 $\Gamma$ point with the three M points in the Brillouin zone (BZ),
 and the vectors $2\,\sB_j$ are primitive reciprocal lattice vectors.
 The bands of this tight-binding model have therefore a higher
 periodicity than the bands of a more general model. This leads to the
 appearance of special symmetry lines (thin lines) and symmetry points
 (M' and K') in the Brillouin zone, shown in Fig.~\ref{zones}, where
 the bands are plotted along the line $\Gamma$'-K'-M'-$\Gamma$'.
 Within a reduced BZ, these bands correspond to the bands of a nearest
 neighbor tight-binding model on a Kagom\'e lattice \cite{maekawa}.
 The density of states per spin and per reduced BZ is also shown in
 Fig.~\ref{zones}. It has a logarithmic singularity at $E=2t$ and
 jumps from $\sqrt{3}/(2\pi t)$ to 0 at $E=4t$.

 The states, that are connected by the considered hopping processes
 form a Kagom\'e lattice.  Since in this way the CoO$_2$-plane consists of
 four independent and inter-penetrating Kagom\'e lattices
 \cite{maekawa}, it is convenient to label the states belonging to the
 same Kagom\'e lattice with an index $l\in\{0,1,2,3\}$. This can be
 done with the vectors $\ba_l$ of Fig.~\ref{CoO} as
 \begin{equation}
   \label{rkagoop}
   a^{\dag l}_{\ssR m}=c^{\dag}_{\ssR+\ba_l+\ba_m m}.
 \end{equation}
 In this way, the operators $a^{\dag l}_{\ssR m}$ with fixed $l$
 create all the states off a Kagom\'e lattice. In the following, these
 operators will be called {\it Kagom\'e operators}.  Their Fourier
 transform is given by
 \begin{equation}
   \label{kkagoop}
   a^{\dag l}_{\ssK m}=\frac{2}{\sqrt{N}}\sum_{\sR} e^{i\ssK\cdot(\ssR+\ba_l+\ba_m)}a^{\dag l}_{\ssR m},
 \end{equation}
 where the vectors $\sK$ belong to the reduced BZ, labeled $0$ in
 Fig.~\ref{zones} and $\sR$ runs over the lattice spanned by the
 vectors $2\ba_i$.
 
 The BZ consists of four reduced BZs shown in Fig.~\ref{zones}. An
 alternative labeling of the states is obtained therefore by defining
 the operators
 \begin{equation}
   \label{kkbgoop}
   b^{\dag j}_{\ssK m}=e^{-i \ssB_j\cdot \ba_m}c^{\dag }_{\ssK+\ssB_j  m},
 \end{equation}
 where the vectors $\sB_j$ are defined in Eq.~(\ref{scalerel}) and in
 addition we set $\sB_0=0$.  As shown in Eq.~(\ref{2defb}) of Appendix
 A, the transformation between the Kagom\'e operators $a^{\dag
   l}_{\ssK}$ and the {\it pocket operators} $b^{\dag j}_{\ssK}$
 corresponds to a discrete Fourier transformation of a $2\times 2$
 lattice, and is given by
 \begin{equation}
   \label{kpocketop}
   b^{\dag j}_{\ssK m}=\frac{1}{2}\sum_{l} e^{i\ssB_j\cdot\ba_l}a^{\dag l}_{\ssK m}=\sum_l\mF_{jl}a^{\dag l}_{\ssK m},
 \end{equation}
 where we have defined the symmetric and orthogonal $4\times 4$ matrix
\begin{equation}
\label{rso}
\mF_{jl}=\mF_{lj}=\mF^*_{jl}=\mF^{-1}_{jl}=\frac{1}{2} e^{i\ssB_j\cdot\ba_l}.
\end{equation} 
Note that the matrix elements of $\mF$ are $\pm 1/2$, as the scalar
products $\sB_j\cdot\ba_l$ of Eq.~(\ref{scalerel}) equal $0$ or $\pm
\pi$.

The tight-binding Hamiltonian (\ref{tbham}) is diagonal in the pocket
indices $j$ (cf. Appendix A Eq.~(\ref{diag})),
 \begin{equation}
 \label{abtbham}
 H_{\mathrm{tb}}=\sum_{l \ssK \s}\sum_{mm'}\epsilon^{mm'}_{\ssK}b^{\dag l}_{\ssK m \sigma}b^l_{\ssK m'\sigma}.
 \end{equation}
 From this expression it is apparent, that the tight-binding
 Hamiltonian is invariant under any $U(4)$ transformation of the of
 the form
 \begin{equation}\label{SU4trans}
 b^{\dag j}_{\ssK m\s}\rightarrow \tilde{b}^{\dag j}_{\ssK m\s}= \sum_{j'} U_{jj'}b^{\dag j'}_{\ssK m \s}.
 \end{equation}
 Eq.~(\ref{kpocketop}) is just a special case of Eq.~(\ref{SU4trans}).
 This shows that $H_{\mathrm{tb}}$ is also diagonal in the Kagom\'e
 indices.
 
 It is important to notice that the transformations in
 Eq.~(\ref{SU4trans}) involves symmetries that are not present in a
 more general tight-binding model.  For example a finite hopping
 integral $t_{dd}$ due to the $\s$-hybridization between neighboring
 $t_{2g}$-orbitals would break this symmetry. We will discuss this
 aspect below in more detail and remain for the time being in this
 high-symmetry situation.
 
 In Na$_x$CoO$_2$ the lower two bands are completely filled and will
 be quite inert.  For this reason in the following sections we will
 only deal with the operators of the top band $E^1_{\bk}$ whose
 operators are denoted as
 \begin{equation}
   \label{shortab}
   a^{\dag l}_{\ssK \s}=\sum_mO^{1m}_{\ssK}a^{\dag l}_{\ssK m\s} \quad\mathrm{and}\quad b^{\dag j}_{\ssK \s}=\sum_mO^{1m}_{\ssK}b^j_{\ssK m\s},
 \end{equation}
 resp., where $O^{1m}_{\ssK}$ are matrix elements of the rotation
 matrix $\hat{O}_{\ssK}$ of Eq.~(\ref{rotmat}).
 
 The top band gives rise to four identical Fermi surface pockets in
 the BZ, one in the $ \Gamma $-point and three at the M-points. A
 translation in the reciprocal space by the vectors $\sB_j$ maps the
 pocket around the $\Gamma$ point onto a pocket around the M-point.
 However, this fact does not lead to nesting singularities in the
 susceptibility because a hole pocket is mapped onto a hole pocket by
 the vector $\sB_j$.  The susceptibility of the top band is given by
\begin{equation}
\label{suscep}
\chi^o_{q}=\frac{1}{N}\sum_k \frac{f_{k+q}-f_{k}}{E^1_k-E^1_{k+q}}=\frac{4}{N}\sum_{\ssK} \frac{f_{\ssK+\ssQ}-f{\ssK}}{E^1_{\ssK}-E^1_{\ssK+\ssQ}},
\end{equation}
where $f_k=f(\beta(E^1_k-\mu))$ and $f$ is the Fermi-function. In the
last expression of Eq.~(\ref{suscep}), the sum over $\sK$ is
restricted to the reduced BZ. $\sQ$ also lies in the reduced BZ and is
given by $\sQ=q+\sB_j$.  The susceptibility $\chi^0_q=\chi^0_{\ssQ}$
is periodic with respect to the reduced BZ and is just four times the
susceptibility of a single Kagom\'e lattice. As we have almost
circular hole pockets with quadratic dispersion around the $\Gamma$
and the M points, the susceptibility is therefore approximately given
by the susceptibility of the free electron gas in two dimensions
within each reduced BZ, with circular plateaus of radius
$2\sK_{\mathrm{F}}$ around the $\Gamma$ and the three M points.

 \section{Coulomb Interaction\label{coulomb}}
 
 In this section we introduce the Coulomb interaction between the
 electrons.  As we have spin and orbital degrees of freedom, the
 on-site Coulomb interaction consists of intra-orbital repulsion $U$,
 inter-orbital repulsion $U'$, Hund's coupling $\Jh$ and a pair
 hopping term $J'$. These parameters are related by $U\approx U'
 +2\Jh$ and $\Jh=J'$, where the first relation is exact for spherical
 symmetry.  We can write the onsite Coulomb interaction as
 \begin{eqnarray}
   \label{inter}
   H^{\mathrm{C}}_{\mathbf{r}}&=& U\sum_{m} n_{\br m\up}n_{\br m \down}
   +
    \frac{U'}{2} \sum_{m\neq m'}\sum_{\s\s'} n_{\br m \s}n_{\br m'\s'}\nn
       &+&\frac{\Jh}{2}\sum_{m\neq m'}\sum_{\s\s'}c^{\dag}_{\br m\s}c^{\dag}_{\br m' \s'}c_{\br m\s'}c_{\br m'\s} \\
 &+&\frac{J'}{2}\sum_{m\neq m'}\sum_{\s\neq \s'}c^{\dag}_{\br m\s}c^{\dag}_{\br m\s'}c_{\br m'\s'}c_{\br m'\s} ,\nonumber
 \end{eqnarray}
 where $n_{\br m \s}=c^{\dag}_{\br m \s}c_{\br m \s}$.  We obtain an
 effective Hamiltonian for the Coulomb interaction by rewriting the
 Hamiltonian in terms of the pocket operators of the top band $b^{\dag
   l}_{\ssK \s}$ defined in Eq.~(\ref{shortab}).  For small
 $\rK=|\sK|a$ we can expand Eq.~(\ref{shortab}) in powers of $\rK^2$
 and obtain up to terms of the order $\rK^2$
 \begin{equation}
   \label{symab}
  b^{\dag j}_{\ssK \s}=\frac{1}{\sqrt{3}}\sum_m \left(1+\frac{\rK^2}{12}\cos{\left[2(\theta-\theta_m)\right]}\right)b^j_{\ssK m\s},
 \end{equation}
 where $\theta_m=2\pi m/3$.  Expanding the energy of the top band
 around the point $\Gamma'$, we obtain
 \begin{equation}\label{enexpa}
 \epsilon^1_\ssK=t(4-\rK^2+\frac{\rK^4}{12}-\frac{\rK^6}{360}\cos(6\theta)+O(\rK^8)).
 \end{equation}
 This shows that the pockets around the points $\Gamma'$ are almost
 perfectly circular. The radius $\rK_{\mathrm{F}}/ a$ of these pockets
 depends on the Na doping $x$. Note that $ x $ corresponds to the
 density of carriers with $ x =1 $ giving a completely filled top
 band. We have $\rK_{\mathrm{F}}^2=\pi(1-x)/\sqrt{3}$.  For the
 interaction in weak-coupling and at low temperatures, the states near
 the Fermi surface are important.  For these states and for not too
 small Na doping $x$ we can neglect the second term in the parenthesis
 of Eq.~(\ref{symab}) compared to 1.  Note, that this condition on $x$
 is not very restrictive. Even for $x=0.35$ the second term together
 with all higher order terms is on the average one order of magnitude
 smaller than 1.  Dropping the second term in Eq.~(\ref{symab})
 spreads the $a_{1g}$ symmetry of the states $b^{\dag j}_{\ssK \s}$,
 which is exact only for $\sK=0$, to all relevant states in the top
 band.  The interaction (\ref{inter}) can now be rewritten in terms of
 the $a_{1g}$ symmetric operators $b^{\dag j}_{\ssK \s}$. Processes
 involving states of the filled lower bands are dropped. The dropping
 of the second term in the parenthesis of Eq.~(\ref{symab}) is a
 considerable simplification because it removes all $\sK$-dependence
 of the potential.

 At this point it is convenient to introduce density and spin density
 operators for the pocket operators of the top band.
 \begin{eqnarray}
   \label{csdoperators}
  \hat{n}^{ij}_{\ssQ}&=&\frac{4}{N}\sum_{\ssK \s}b^{\dag
     i}_{\ssK+\ssQ \s}b^{j}_{\ssK \s}\\
  \hat{\mathbf{S}}^{ij}_{\ssQ}&=&\frac{2}{N}\sum_{\ssK \s\s'}b^{\dag
     i}_{\ssK+\ssQ\s}\boldsymbol{\s}_{\s\s'}b^{j}_{\ssK \s}\nonumber
 \end{eqnarray}
 The resulting effective interaction can be expressed with these
 operators in the following way
 \begin{equation}
   \label{charsp}
 H_{\mathrm{eff}}=\frac{N}{32}\sum_{\ssQ}\left(B^{\mathrm{s}}_{ijkl} \hat{\mathbf{S}}^{ij}_{\ssQ} \hat{\mathbf{S}}^{lk}_{-\ssQ} +\frac{1}{4} B^{\mathrm{c}}_{ijkl} \hat{n}^{ij}_{\ssQ}
   \hat{n}^{lk}_{-\ssQ}\right).
 \end{equation}
 The symbols $B^{\mathrm{c/s}}$ depend on the Coulomb integrals and
 are given by
 \begin{eqnarray}
 \label{McMs}
 B^{\mathrm{c/s}}_{ijkl}&=&\pm C(2\delta_{ijkl}-\epsilon^2_{ijkl})\pm D\delta_{il}\delta_{jk}\\
 &&+E^{\mathrm{c/s}}\delta_{ij}\delta_{kl}+F^{\mathrm{c/s}}\delta_{ik}\delta_{jl},\nonumber
 \end{eqnarray}
 where the $\delta$ ($\epsilon^2$) symbol equals 1, if all the indices
 are equal (different) and 0 otherwise.  The coefficients $C$, $D$,
 $E^{\mathrm{c/s}}$ and $F^{\mathrm{c/s}}$ are listed in
 TABLE~\ref{coeff}.
 \begin{table}
   \begin{center}
     $\begin{array}{l|l} \hline\hline {9C =\! -3U\!+\!2J'\!+\!2\Jh
         \!+\!2U'}& {9D =\!+3U\!+\!6J'\!-\!2\Jh\!-\!2U'}\\\hline
       {9E^{\mathrm{c}}\!=\! +3U\!-\!2J'\!-\!10\Jh \!+\!14U'}&
       {9E^{\mathrm{s}}\!=\! -3U\!+\!2J'\!-\!6\Jh+2U'}\\\hline
       {9F^{\mathrm{c}}\!=\! +3U\!-\!2J'\!+\!14\Jh \!-\!10U'}&
       {9F^{\mathrm{s}}\!=\! -3U\!+\!2J'\!+\!2\Jh\!-\!6U'}\\
       \hline\hline
     \end{array}$
   \end{center}
 \caption{\label{coeff} The coefficients of Eq.~(\ref{McMs})}
 \end{table}
 Note, that for small pockets, the momenta $\sK$ of the pocket
 operators $b^j_{\ssK}$ in the four fermion terms of Eq~(\ref{charsp})
 can not add up to a half a reciprocal lattice vector $\sB_i$. In
 order to conserve momentum they must therefore add up to zero.  Due
 to the position of the pockets in the BZ, Umklapp processes with low
 energy transfer are however possible for arbitrary small pockets. In
 fact, the processes proportional to $\epsilon^2_{ijkl}$ and
 $\delta_{il}\delta_{jk}(1-\delta_{ij})$ are Umklapp processes, as
 $\sB_i-\sB_j+\sB_l-\sB_k$ is a non-vanishing reciprocal lattice
 vector for $\epsilon_{ijkl}\neq 0$ and for
 $\delta_{il}\delta_{jk}(1-\delta_{ij})\neq 0$, and from
 Eq.~(\ref{kkbgoop}) the momentum created by the operator $b^{\dag
   j}_{\ssK}$ is $\sK+\sB_j$.
 
 Some details about the derivation of Eq.~(\ref{charsp}) are provided
 in Appendix \ref{deriv}. There are different ways of writing this
 interaction in terms of the operators in ($\ref{csdoperators}$). Our
 formulation treats charge- and spin degrees of freedom in the same
 way. It corresponds to the decomposition of a Hubbard interaction
 $n_{\up}n_{\down}$ into
 $\frac{1}{2}(\frac{1}{4}n^2-\mathbf{S}\cdot\mathbf{S})$.
 
 In order to express the effective interaction Hamiltonian of
 Eq.~(\ref{charsp}) in terms of the Kagom\'e operators $a^l_{\ssK
   \s}$, we define spin and charge density operators from the Kagom\'e
 operators $a^{l}_{\ssK \s}$ as in Eq.~(\ref{csdoperators}).
 \begin{eqnarray}
   \label{acsdoperators}
  n^{ij}_{\ssQ}&=&\frac{4}{N}\sum_{\ssK \s}a^{\dag
     i}_{\ssK+\ssQ \s}a^{j}_{\ssK \s}\\
  \mathbf{S}^{ij}_{\ssQ}&=&\frac{2}{N}\sum_{\ssK \s\s'}a^{\dag
     i}_{\ssK+\ssQ\s}\boldsymbol{\s}_{\s\s'}a^{j}_{\ssK \s}\nonumber
 \end{eqnarray}
 Note, that the density operators, which are defined from the pocket
 operators $b^j_{\ssK \sigma}$ are marked by a hut. The effective
 Hamiltonian, $H_{\mathrm{eff}}$, of Eq.~(\ref{charsp}) can be
 rewritten as
 \begin{equation}
   \label{charspA}
 H_{\mathrm{eff}}=\frac{N}{32}\sum_{\ssQ}\left(A^{\mathrm{s}}_{ijkl} \mathbf{S}^{ij}_{\ssQ} \mathbf{S}^{lk}_{-\ssQ} +\frac{1}{4} A^{\mathrm{c}}_{ijkl} n^{ij}_{\ssQ}
   n^{lk}_{-\ssQ}\right).
 \end{equation}
 From Eq.~(\ref{kpocketop}) and (\ref{rso}) follows that
 \begin{equation}
 A^{c/s}_{ijkl}=\mF_{im}\mF_{jn}\mF_{ko}\mF_{lp}B^{c/s}_{mnop}.
 \end{equation}
 The symbols $A^{c/s}$ turn out to have a simpler structure, given by
 \begin{eqnarray}
   \label{simpleA}
 A^{c}_{ijkl}&=&\frac{8}{9}\bigg[-\frac{C}{2}\delta_{ijkl}+J'\delta_{il}\delta_{jk}+(2U'-\Jh)\delta_{ij}\delta_{kl}+{}\nn
 &&\quad\;{}+(2\Jh-U')\delta_{ik}\delta_{jl}\bigg],\\
 A^{s}_{ijkl}&=&\frac{8}{9}\bigg[+\frac{C}{2}\delta_{ijkl}-J'\delta_{il}\delta_{jk}-\Jh\delta_{ij}\delta_{kl}-U'\delta_{ik}\delta_{jl}\bigg].\nonumber
 \end{eqnarray}

 \section{$\mathbf{SU(4)}$ generators\label{SU4order}}
 The tight-binding Hamiltonian described in section
 \ref{tight-binding} has a $U(4)$ symmetry, reflecting the fact that
 it consists of 4 independent and equivalent Kagom\'e lattices. The
 correlations introduced by the on-site Coulomb repulsion in
 Eq.~(\ref{inter}) breaks this symmetry and leads to interaction
 between orbitals belonging to different Kagom\'e lattices, as the
 three $t_{2g}$-orbitals on a given Co-site belong to three different
 Kagom\'e lattices.  The effective Hamiltonian in Eq.~(\ref{charsp})
 is not invariant under general $U(4)$ transformations, but
 is still invariant under a finite subgroup of $U(4)$.  The symbols
 $A^{c/s}_{ijkl}$ defined in Eq.~(\ref{simpleA}) are invariant under
 permutation of the indices, i.e.\ 
 \begin{equation}
 \label{Mcspert}
 A^{s/c}_{ijkl}=A^{s/c}_{\mP(i)\mP(j)\mP(k)\mP(l)}\qquad \mP\in\mS_4.
 \end{equation}
 From this follows that the symmetric group $\mS_4$ is a subgroup of
 $G$.  Multiplying all operators $a^l_{\ssK,\sigma}$ with the same
 Kagom\'e index $l$ by $-1$ also leaves the Hamiltonian,
 $H_{\mathrm{eff}}$, invariant, because the symbols $A^{c/s}_{ijkl}$
 are nonzero only if the four indices $ijkl$ are pairwise equal. These
 two different symmetry operations generate a group with $384$
 elements. This group $G$ is isomorphic to the symmetry group of the
 four-dimensional hypercube.  In appendix~\ref{QandG} the structure of
 the group $G$ is discussed and a character table is shown.
 
 To proceed, let $Q^r$ $r=0,\dots,15$ be a basis in the 16 dimensional
 real vector space, $V$ of Hermitian $4\times 4$ matrices, fulfilling
 the usual orthonormality and completeness relations
 \begin{equation}
   \label{orthogore}
   Q^r_{ij} Q^l_{ji}=\frac{1}{2}\delta_{rl}\qquad \sum_{r=0}^{15}Q^r_{ij}Q^r_{kl}=\frac{1}{2}\delta_{il}\delta_{jk}.
 \end{equation}
 This basis can be chosen such, that $Q^0$ is proportional to the unit
 matrix, $Q^{1-3}$ are diagonal, $Q^{4-9}$ are real and $Q^{10-15}$
 are imaginary.  It is convenient to define also the {\it dual}
 matrices
\begin{equation}
\label{dualmat}
K^r_{ij}=\mF_{im}\mF_{jn}Q^r_{mn}.
\end{equation}  
In TABLE~\ref{Q1-Q15} a choice of a basis $Q^r$, which is particularly
suitable for our purposes, is shown together with the dual basis
$K^r$.  A representation $\rho$ of the group $G$ on $V$ is given by
$\rho(g)Q^r=N^{\top}_gQ^rN_g$ for $g\in G$, where $N_g$ is the natural
four-dimensional representation of $G$ (cf.\ appendix~\ref{QandG}).
The representation $\rho$ is reducible and $V$ is the direct sum of
the four irreducible subspaces $V^0$, $V^{1-3}$, $V^{4-9}$ and
$V^{10-15}$ spanned by matrices $Q^0$, $Q^{1-3}$, $Q^{4-9}$ and
$Q^{10-15}$, respectively.  Therefore, the chosen Basis is appropriate
for the symmetry group $G$.  Defining charge and spin density
operators
 \begin{eqnarray}
   \label{pers}
 n^{r}_{\ssQ}&=&Q^r_{ij}n^{ij}_{\ssQ}=K^r_{ij}\hat{n}^{ij}_{\ssQ}\\ 
\mathbf{S}^{r}_{\ssQ}&=&Q^r_{ij}\mathbf{S}^{ij}_{\ssQ}=K^r_{ij}\hat{\mathbf{S}}^{ij}_{\ssQ}\nonumber,
 \end{eqnarray}
 the interaction Hamiltonian can be written in a diagonal form as
 \begin{equation}
   \label{diagchsp}
  H_{\mathrm{eff}}=\frac{N}{8}\sum_{r=0}^{15}\sum_{\ssQ}\left(\Lambda^{\mathrm{s}}_{r}\mathbf{S}^{r}_{\ssQ} \mathbf{S}^{r}_{-\ssQ} +\frac{1}{4}\Lambda^{\mathrm{c}}_{r}n^{r}_{\ssQ}n^{r}_{-\ssQ}\right).
 \end{equation}
 The coupling constants $\Lambda^{\mathrm{c/s}}_{r}$ are equal for all
 $Q^r$ belonging to the same irreducible subspace in $V$.  They are
 given in TABLE~\ref{Gaal}.
 \begin{table*}
 \begin{ruledtabular}
 \begin{tabular}{rrrrr}$
 \rule{0cm}{0.5cm}Q^1\ (\Gamma^a_5)\hspace{0.3cm}$&$Q^4\ (\Gamma^b_1)\hspace{0.3cm}$&$Q^7\ (\Gamma^b_5)\hspace{0.3cm}$&$Q^{10}\ (\Gamma_4)\hspace{0.2cm}$&$Q^{13}\ (\Gamma_5^c)
 \hspace{0.2cm}$\\$
 \frac{1}{2\sqrt{2}}\left(\begin{matrix} 1 & 0 & 0 & 0 \cr 0 & 1 & 0 & 0 \cr 0 & 0 & \bar{1} & 0 \cr 0 & 0 & 0 & \bar{1} \cr  \end{matrix}\right)
 $&$
 \rule{0cm}{1.1cm}\frac{1}{2\sqrt{6}}\left(\begin{matrix} 0 & 1 & 1 & 1 \cr 1 & 0 & 1 & 1 \cr 1 & 1 & 0 & 1 \cr 1 & 1 & 1 & 0 \cr  \end{matrix}\right)
 $&$
 \frac{1}{2\sqrt{2}}\left(\begin{matrix} 0 & 1 & 0 & 0 \cr 1 & 0 & 0 & 0 \cr 0 & 0 & 0 & \bar{1} \cr 0 & 0 & \bar{1} & 0 \cr  \end{matrix}\right)
 $&$
 \frac{1}{4}\left(\begin{matrix} 0 & 0 & \bar{i} & i \cr 0 & 0 & i & \bar{i} \cr i & \bar{i} & 0 & 0 \cr \bar{i} & i & 0 & 0 \cr  \end{matrix}\right)
 $&$
 \frac{1}{4}\left(\begin{matrix} 0 & 0 & \bar{i} & \bar{i} \cr 0 & 0 & \bar{i} & \bar{i} \cr i & i & 0 & 0 \cr i & i & 0 & 0 \cr \end{matrix}\right)
 $\\
 $\rule{0cm}{0.5cm}Q^2\ (\Gamma^a_5)\hspace{0.3cm}$&$Q^5\ (\Gamma_3)\hspace{0.3cm}$&$Q^8\ (\Gamma^b_5)\hspace{0.3cm}$&$Q^{11}\ (\Gamma_4)\hspace{0.2cm}$&$Q^{14}\ (\Gamma^c_5)\hspace{0.2cm}$\\$
 \rule{0cm}{1.1cm}\frac{1}{2\sqrt{2}}\left(\begin{matrix} 1 & 0 & 0 & 0 \cr 0 & \bar{1} & 0 & 0 \cr  0  & 0 & 1 & 0 \cr 0 &  0 & 0 & \bar{1} \cr  \end{matrix}\right)
 $&$
 \frac{1}{4}\left(\begin{matrix} 0 & 1 & \bar{1} & 0 \cr 1 & 0 & 0 & \bar{1} \cr \bar{1} & 0 & 0 & 1 \cr 0 & \bar{1} & 1 & 0 \cr \end{matrix}\right)
 $&$
 \frac{1}{2\sqrt{2}}\left(\begin{matrix} 0 & 0 & 1 & 0 \cr 0 & 0 & 0 & \bar{1} \cr 1 & 0 & 0
     & 0 \cr 0 & \bar{1} & 0 & 0 \cr  \end{matrix}\right)
 $&$
 \frac{1}{4}\left(\begin{matrix} 0 & i & 0 & \bar{i} \cr \bar{i} & 0 & i & 0 \cr 0 & \bar{i} & 0 & i \cr i & 0 & \bar{i} & 0 \cr \end{matrix}\right)
 $&$
 \frac{1}{4}\left(\begin{matrix} 0 & \bar{i} & 0 & \bar{i} \cr i & 0 & i & 0 \cr 0 & \bar{i} & 0 & \bar{i} \cr i & 0 & i & 0 \cr \end{matrix}\right)
 $\\
 $\rule{0cm}{0.5cm}Q^3\ (\Gamma^a_5)\hspace{0.3cm}$&$Q^6\ (\Gamma_3)\hspace{0.3cm}$&$Q^9\ (\Gamma^b_5)\hspace{0.3cm}$&$Q^{12}\ (\Gamma_4)\hspace{0.2cm}$&$Q^{15}\ (\Gamma_5^c)\hspace{0.2cm}$\\$
 \rule{0cm}{1.1cm}\frac{1}{2\sqrt{2}}\left(\begin{matrix} 1 & 0 & 0 & 0 \cr 0 & \bar{1} & 0 & 0 \cr 0 & 0 & \bar{1} & 0 \cr 0 & 0 & 0 & 1 \cr \end{matrix}\right)
 $&$
 \frac{1}{4\sqrt{3}}\left(\begin{matrix} 0 & 1 & 1 & \bar{2} \cr 1 & 0 & \bar{2} & 1 \cr 1 & \bar{2} & 0 & 1 \cr \bar{2} & 1 & 1 & 0 \cr \end{matrix}\right)
 $&$
 \frac{1}{2\sqrt{2}}\left(\begin{matrix} 0 & 0 & 0 & 1 \cr 0 & 0 & \bar{1} & 0 \cr 0 & \bar{1} & 0 & 0 \cr 1 & 0 & 0 & 0 \cr  \end{matrix}\right)
 $&$
 \frac{1}{4}\left(\begin{matrix} 0 & \bar{i} & i & 0 \cr i & 0 & 0 & \bar{i} \cr \bar{i} & 0 & 0 & i \cr 0 & i & \bar{i} & 0 \cr  \end{matrix}\right)
 $&$
 \frac{1}{4}\left(\begin{matrix} 0 & \bar{i} & \bar{i} & 0 \cr i & 0 & 0 & i \cr i & 0 & 0 & i \cr 0 & \bar{i} & \bar{i} & 0 \cr \end{matrix}\right)
 $\\
 \rule{0cm}{0.1cm}\\
 \hline
$ \rule{0cm}{0.5cm}K^1\ (\Gamma^a_5)\hspace{0.3cm}$&$K^4\ (\Gamma^b_1)\hspace{0.3cm}$&$K^7\ (\Gamma^b_5)\hspace{0.3cm}$&$K^{10}\ (\Gamma_4)\hspace{0.2cm}$&$K^{13}\ (\Gamma_5^c)
 \hspace{0.2cm}$\\$
 \frac{1}{2\sqrt{2}}\left(\begin{matrix} 0 & 1 & 0 & 0 \cr 1 & 0 & 0 & 0 \cr 0 & 0 & 0 & 1 \cr 0 & 0 & 1 & 0 \cr  \end{matrix}\right)
 $&$
 \rule{0cm}{1.1cm}\frac{1}{2\sqrt{6}}\left(\begin{matrix} 3 & 0 & 0 & 0 \cr 0 & \bar{1} & 0 & 0 \cr 0 & 0 & \bar{1} & 0 \cr 0 & 0 & 0 & \bar{1} \cr  \end{matrix}\right)
 $&$
 \frac{1}{2\sqrt{2}}\left(\begin{matrix} 0 & 1 & 0 & 0 \cr 1 & 0 & 0 & 0 \cr 0 & 0 & 0 & \bar{1} \cr 0 & 0 & \bar{1} & 0 \cr  \end{matrix}\right)
 $&$
 \frac{1}{2}\left(\begin{matrix} 0 & 0 & 0 & 0 \cr 0 & 0 & 0 & 0 \cr 0 & 0 & 0 & i \cr 0 & 0 & \bar{i} & 0 \cr  \end{matrix}\right)
 $&$
 \frac{1}{2}\left(\begin{matrix} 0 & i & 0 & 0 \cr \bar{i} & 0 & 0 & 0 \cr 0 & 0 & 0 & 0 \cr 0 & 0 & 0 & 0 \cr \end{matrix}\right)
 $\\
 $\rule{0cm}{0.5cm}K^2\ (\Gamma^a_5)\hspace{0.3cm}$&$K^5\ (\Gamma_3)\hspace{0.3cm}$&$K^8\ (\Gamma^b_5)\hspace{0.3cm}$&$K^{11}\ (\Gamma_4)\hspace{0.2cm}$&$K^{14}\ (\Gamma_5^c)\hspace{0.2cm}$\\$
 \rule{0cm}{1.1cm}\frac{1}{2\sqrt{2}}\left(\begin{matrix} 0 & 0 & 1 & 0 \cr 0 & 0 & 0 & 1 \cr  1  & 0 & 0 & 0 \cr 0 &  1 & 0 & 0 \cr  \end{matrix}\right)
 $&$
 \frac{1}{2}\left(\begin{matrix} 0 & 0 & 0 & 0 \cr 0 & 1 & 0 & 0 \cr 0 & 0 & \bar{1} & 0 \cr 0 & 0 & 0 & 0 \cr \end{matrix}\right)
 $&$
 \frac{1}{2\sqrt{2}}\left(\begin{matrix} 0 & 0 & 1 & 0 \cr 0 & 0 & 0 & \bar{1} \cr 1 & 0 & 0
     & 0 \cr 0 & \bar{1} & 0 & 0 \cr  \end{matrix}\right)
 $&$
 \frac{1}{2}\left(\begin{matrix} 0 & 0 & 0 & 0 \cr 0 & 0 & 0 & \bar{i} \cr 0 & 0 & 0 & 0 \cr 0 & i & 0 & 0 \cr \end{matrix}\right)
 $&$
 \frac{1}{2}\left(\begin{matrix} 0 & 0 & i & 0 \cr 0 & 0 & 0 & 0 \cr \bar{i} & 0 & 0 & 0 \cr 0 & 0 & 0 & 0 \cr \end{matrix}\right)
 $\\
 $\rule{0cm}{0.5cm}K^3\ (\Gamma^a_5)\hspace{0.3cm}$&$K^6\ (\Gamma_3)\hspace{0.3cm}$&$K^9\ (\Gamma^b_5)\hspace{0.3cm}$&$K^{12}\ (\Gamma_4)\hspace{0.2cm}$&$K^{15}\ (\Gamma_5^c)\hspace{0.2cm}$\\$
 \rule{0cm}{1.1cm}\frac{1}{2\sqrt{2}}\left(\begin{matrix} 0 & 0 & 0 & 1 \cr 0 & 0 & 1 & 0 \cr 0 & 1 & 0 & 0 \cr 1 & 0 & 0 & 0 \cr \end{matrix}\right)
 $&$
 \frac{1}{2\sqrt{3}}\left(\begin{matrix} 0 & 0 & 0 & 0 \cr 0 & 1 & 0 & 0 \cr 0 & 0 & 1 & 0 \cr 0 & 0 & 0 & \bar{2} \cr \end{matrix}\right)
 $&$
 \frac{1}{2\sqrt{2}}\left(\begin{matrix} 0 & 0 & 0 & 1 \cr 0 & 0 & \bar{1} & 0 \cr 0 & \bar{1} & 0 & 0 \cr 1 & 0 & 0 & 0 \cr  \end{matrix}\right)
 $&$
 \frac{1}{2}\left(\begin{matrix} 0 & 0 & 0 & 0 \cr 0 & 0 & i & 0 \cr 0 & \bar{i} & 0 & 0 \cr 0 & 0 & 0 & 0 \cr  \end{matrix}\right)
 $&$
 \frac{1}{2}\left(\begin{matrix} 0 & 0 & 0 & i \cr 0 & 0 & 0 & 0 \cr 0 & 0 & 0 & 0 \cr \bar{i} & 0 & 0 & 0 \cr \end{matrix}\right)
 $
 \end{tabular}
 \end{ruledtabular}
 \caption{\label{Q1-Q15}The matrices $Q^{1-15}$ are a choice of an orthonormal complete basis of the 15 dimensional real vector space of traceless hermitian matrices, so called generators of $SU(4)$, that is adequate to the symmetry of the CoO$_2$-layer. The matrices $K^r$ are obtained from $Q^r$ by Eq.~(\ref{dualmat}).  Note that $\bar{1}=-1$ and $\bar{i}=-i$. $2\sqrt{2}Q^0=2\sqrt{2}K^0$ is the $4\times4$ unit matrix.}
 \end{table*}
 \begin{table*}[t]
 \begin{ruledtabular}
 \begin{tabular}{ccccc}
 $r$ & 0 & 1--3 & 4--9 & 10--15 \\
 \hline
 $\Lambda^{\mathrm{c}}_r$ & $\frac{2}{9}(3U+12U'-6\Jh)$  & $\frac{2}{9}(3U-4U'+2\Jh)$ & $\frac{2}{9}(-2U'+4\Jh-2J')$ & $\frac{2}{9}(-2U'+4\Jh+2J')$\\
 $\Lambda^{\mathrm{s}}_r$ & $-\frac{2}{9}(3U+6\Jh)$ & $-\frac{2}{9}(3U-2\Jh)$ & $-\frac{2}{9}(2U'-2J')$ & $-\frac{2}{9}(2U'+2J')$
 \end{tabular}
 \end{ruledtabular}
 \caption{\label{Gaal} The coefficients $\Lambda^{\mathrm{s/c}}_r$.}
 \end{table*}
 \section{Reduction of the symmetry\label{breaksym}}
 The tight-binding Hamiltonian in Eq.~(\ref{abtbham}) has a $U(4)$
 symmetry and even after introducing Coulomb interaction, the
 effective Hamiltonian (\ref{diagchsp}) is invariant under the
 symmetry group $G$. In a real CoO$_2$ plane this symmetry is reduced even
 in the paramagnetic state. There are terms in the Hamiltonian of the
 real system that restrict the symmetry operations of $G$ to the
 subgroup, which describes real crystallographic space-group
 symmetries.
 
 A trigonal distortion of the oxygen octahedra by approaching the two
 O-layers to the Co-layer, is for example compatible with the point
 group symmetry $D_{3d}$ of the CoO$_2$ layer. However it lifts the
 degeneracy of the $t_{2g}$-orbitals, leading to a term
\begin{eqnarray}
\label{trigo}
H_{\mathrm{tr}}&=&D_{\mathrm{tr}}\sum_{k\sigma}\sum_{m\neq m'}c^{\dag}_{k m\sigma}c_{k m' \sigma}\\
&=&D_{\mathrm{tr}}\sum_{l \ssK\sigma}\sum_{m\neq m'}b^{\dag l}_{\ssK m\sigma}b^l_{\ssK m' \sigma}e^{i\ssB_l\cdot (\mathbf{a}_m-\mathbf{a}_{m'})}\nonumber
\end{eqnarray}
in the Hamiltonian, where we used Eq.~(\ref{kkbgoop}) to obtain the
second line.  For the top band we obtain with Eq.~(\ref{symab}) and
(\ref{summmp})
\begin{eqnarray}
\label{toptrigo}
H_{\mathrm{tr}}&=&\sqrt{2/3}\;D_{\mathrm{tr}}\;4\sum_{l\ssK\sigma}(K^4_{ll} +O(\rK^2))\;b^{\dag l}_{\ssK\sigma}b^l_{\ssK \sigma}\nonumber\\
&\approx&\sqrt{2/3}\;D_{\mathrm{tr}}N n^4_{\scriptscriptstyle{\mathbf{0}}}
\end{eqnarray}
where the matrix $K^4$ is given in TABLE~\ref{Q1-Q15} and $\rK=|\sK|
a$ is small for the relevant states near the Fermi pockets, if the
pockets are small enough.
Similarly, a finite direct hopping integral $t_{dd}$ leads to the term
\begin{eqnarray}
\label{direct}
H_{dd}&=& t_{dd}\sum_{km\sigma}2\cos{(k_m)}c^{\dag}_{km\sigma}c_{km\sigma}\nonumber\\
&= &4\sqrt{6}t_{dd}\sum_{l\ssK\sigma}(K^4_{ll}+O(\rK^2))b^{\dag l}_{\ssK\sigma}b^l_{\ssK\sigma}\nonumber\\
&\approx &\sqrt{6}t_{dd} N n^4_{\scriptscriptstyle{\mathbf{0}}}
\end{eqnarray}
where we again dropped the terms involving the lower bands in the
second line.  In fact, any other additional hopping term or any
quadratic perturbation compatible with the space group is proportional
to the field $n^4_{\scriptscriptstyle{\mathbf{0}}}$ in the limit of
small pockets, if the perturbation is diagonal in the spin indices.
As the trigonal distortion of the octahedra is nonzero and additional
hopping terms are present in the CoO$_2$-layer, a term proportional to
$n^4_{\scriptscriptstyle{\mathbf{0}}}$ exists in the Hamiltonian
acting like a symmetry breaking field. For simplicity, we will refer
to a term proportional to $n^4_{\scriptscriptstyle{\mathbf{0}}}$ in
the Hamiltonian as the {\em trigonal distortion}, even though this
term is rather an effective trigonal distortion that also includes the
effects of additional hopping terms.
 
From the matrix $K^4$ can be seen, that the presence of a finite
field, $n^4_{\scriptscriptstyle{\mathbf{0}}}$, in the Hamiltonian
leads to a distinction between the $\Gamma$ and the M points in the BZ
and the four hole pockets are no longer equivalent.  In real space,
the four Kagom\'e lattices are still equivalent, as they transform
under space group symmetries among themselves.  In fact, the matrix
$Q^4$ is still invariant under permutations of rows and columns, i.e.\ 
$N^T_gQ^4N_g=Q^4$ for all $g\in\mS_4$, but $Q^4$ is not invariant
under changing the sign of all operators with the same Kagom\'e index.
These sign changes, however, are not space-group symmetries, but gauge
symmetries, originating from the fact that the charge on the Kagom\'e
lattices is conserved by $H_{\mathrm{tb}}$ and also by the Coulomb
interaction except for the pair-hopping term proportional to $J'$ in
Eq.~(\ref{inter}). This term however can only change the number of
electrons by two, leading to these gauge symmetries, that are broken,
as soon as single electron hopping processes between the Kagom\'e
lattices are introduced.

To classify the states according to the real symmetry group of the
CoO$_2$-layer without gauge symmetries, it is therefore sufficient, to
consider the presence of a small field
$n^4_{\scriptscriptstyle{\mathbf{0}}}$, that restricts the symmetry
group $G$ to a subgroup, consisting of space group symmetries of the
CoO$_2$-layer. This subgroup of $G$ is isomorphic to $\mS_4\simeq
T_d\simeq O$.  Intuitively it is understandable that the symmetry of
the four dimensional cube reduces to the symmetry of a three
dimensional cube, if one of the four hole pockets is not equivalent to
the other three.

Form TABLE~\ref{Q1-Q15} can be seen, that the matrices $Q^0$,
$Q^{1-3}$, $Q^4$, $Q^{5-6}$, $Q^{7-9}$, $Q^{10-12}$ and $Q^{13-15}$
transform irreducibly under $\mS_4$ with the representations
$\Gamma^a_{1}$, $\Gamma^a_{5}$, $\Gamma^b_{1}$, $\Gamma_3$,
$\Gamma^b_{5}$, $\Gamma_4$ and $\Gamma_5^c$, respectively, where the
upper-script letter distinguishes between different subspaces
transforming with the same representation.
 
The appearance of three dimensional irreducible representations in the
classification of the order parameters can be understood as follows.
The point group $P$ of a single CoO$_2$-layer is $D_{3d}$, and the degree
of its irreducible representations is $\le 2$.  The point group is the
factor group $S/T$ where $S$ is the space group of the CoO$_2$-layer and
$T$ is the subgroup of all pure translations. For our system it is
convenient to consider the factor group $P'=S/2T$, where $2T$ is the
subgroup of $T$ that is generated by translations of $2\mathbf{a}_i$.
$P'$ is isomorphic to the cubic group $O_h$ and has irreducible
representations of degree 3.  The operators $n^r_{\ssQ}$ and
$\mathbf{S}^r_{\ssQ}$ transforms irreducibly under the translations in
$2T$ for every $r$.  The symmetry operations of $P'$ however mix
operators $n^r_{\ssQ}$ (or $\mathbf{S}^r_{\ssQ}$) with different $r$,
and the irreducible representations as given above or shown in
TABLE~\ref{Q1-Q15} are obtained.  Strictly speaking, the basis of
SU(4) generators shown in TABLE~\ref{Q1-Q15} is the correct eigenbasis
only for an infinitesimal small trigonal distortion, for a finite
distortion, the representations $\Gamma_1^a$ and $\Gamma_1^b$ as well
as $\Gamma_5^a$ and $\Gamma_5^b$ can hybridize as they transform with
the same irreducible representation.  Note, that $\Gamma_5^c$
transforms differently under time reversal.  The situation here is
similar to atomic physics, where a crossover from the Zeeman effect to
the Paschen-Back effect with increasing magnetic field occurs, because
states with the same $J_z$ can hybridize.

 \section{Ordering Patterns\label{OP}}
 
 In this section the different types of symmetry breaking phase
 transitions are discussed in a mean-field picture. The symmetry
 breaking is due to existence of a finite order parameter, that is in
 our case given by the expectation value $\langle n^r_{\ssQ}\rangle$
 of $\langle \mathbf{S}^r_{\ssQ}\rangle$. Note, that a finite
 expectation value $\langle
 n^0_{\scriptscriptstyle{\mathbf{0}}}\rangle$ or $\langle
 n^4_{\scriptscriptstyle{\mathbf{0}}}\rangle$ does not break any
 symmetry of the CoO$_2$-layer.
 
 In our tight-binding model as it was discussed in section
 \ref{tight-binding}, the susceptibility, $\chi^0$ is given by 4
 identical plateaux around the $\Gamma$ and the M points.  In the
 presence of a trigonal distortion, the susceptibility still keeps a
 plateaux like structure but the diameter of the plateaux decreases,
 such that the susceptibility appears sharply enhanced around the M and
 the $\Gamma$ points.  Therefore we restrict the discussion to the
 case where $\sQ$ equals zero and write $n^r$ and $\mathbf{S}^r$
 instead of $n^r_{\scriptscriptstyle{\mathbf{0}}}$ and
 $\mathbf{S}^r_{\scriptscriptstyle{\mathbf{0}}}$ from now on.  Note,
 that in our formalism the states with $\sQ=0$ describe periodic
 states with the enlarged unitcell of the Kagom\'e lattice. But the
 internal degrees of freedom within this enlarged unitcell still
 allows for rather complicated charge- and spin-patterns.  States with
 a small but finite $\sQ$ describe modulations of these local states
 on long wavelengths.  It is therefore important to understand first
 the local states the are described by $\sQ=0$ instabilities.
 Furthermore, only $\sQ=0$ states couple to the periodic potential
 produced by a Na superstructure at $x=0.5$.

 The $\sQ=0$ instabilities lead to a chemical potential difference for
 states belonging to different hole pockets.  In general, the BZ is
 folded and states of different hole pockets combine to new
 quasi-particles.  In this case, translational and/or rotational
 symmetry is broken.  Complex ordering patterns can be realized
 without opening of gaps, i.e.\ the system stays metallic.
 
 We consider first the orderings given by a finite expectation value
 of the charge density operators $n^r$.  This expectation value is
 given by
\begin{equation}
\label{modes}
\langle n^r\rangle= \frac{4}{N}\sum_{\ssK \sigma}\lambda^r_l \langle v^{\dag l}_{\ssK\sigma}v^l_{\ssK\sigma}\rangle,
\end{equation}
where $\lambda^r_l$ are the eigenvalues of the matrix $Q^r$
($U^r_{ki}Q^r_{ij}\bar{U}^r_{lj}=\lambda^r_l\delta_{kl}$) and
$v^{l}_{\ssK\sigma}=U^r_{ln}a^{n}_{\ssK \sigma}$ are the creation
operators of the quasi-particles.  If only one $\langle n^r
\rangle\neq 0$, the effective interaction Hamiltonian in the
mean-field approximation reduces to
\begin{equation}
\label{meanmodes}
\frac{\Lambda^{\mathrm{c}}_r\langle n^r \rangle}{4} \sum_{\ssK \sigma}\lambda^r_l  v^{\dag l}_{\ssK\sigma}v^l_{\ssK\sigma}.
\end{equation}
If the coupling constant $\Lambda^{\mathrm{c}}_r$ is negative, the
interaction energy of the system can be lowered by introducing an
imbalance between the occupation numbers $n_l=\sum_{\ssK
  \sigma}\langle v^{\dag l}_{\ssK\sigma}v^l_{\ssK\sigma}\rangle $.
The operators $v^{\dag}_{\ssK s}$ create Bloch states with momentum
$\sK$ in the reduced BZ. The amplitudes of the three $t_{2g}$ orbitals
on a given Co site with these Bloch states can be obtained from
Eq.~(\ref{rkagoop}) and (\ref{kkagoop}) and the relation $a^{\dag
  l}_{\ssK}\approx 1/\sqrt{3}\sum_m a^{\dag l}_{\ssK m}$ which follows
from Eq.~(\ref{kpocketop}) and (\ref{symab}).

For the matrices $Q^{0-4}$, these Bloch states are given by a single
$t_{2g}$ orbital on each Co site.  For the non-diagonal matrices
$Q^{4-9}$ these Bloch states are on each Co site proportional to a
linear combination of $t_{2g}$ orbitals of the form
\begin{equation}
  \label{eq:raumdiag}
\frac{1}{\sqrt{3}}(  s_x d_x+s_y d_y+s_z d_z )\ \ \textrm{with}\ \  s_x,s_y,s_z\in\{\pm1\}.
\end{equation}
This linear combination is the atomic $d$-orbital
$\varphi_{0}\equiv\mathcal{Y}_{20}$ parallel to the body-diagonal
$[s_x,s_y,s_z]$ of the cubic unit-cell around a Co atom.

The eigenvectors of the of the matrices $Q^{10-15}$ are complex. A
complex linear combination of $t_{2g}$ orbitals has in general a
non-vanishing expectation value of the orbital angular momentum
operator $\bf{L}$.  In TABLE~\ref{lincom} the angular momentum
expectation values, which are relevant for our discussion are shown.
\begin{table}[htbp]
\begin{center}
\begin{tabular}{llr}
\hline 
\hline
$(id_x+d_y+d_z)/\sqrt{3} $&$\langle \mathbf{L} \rangle =\hbar (0,-1,1)2/3\ $ & (cyclic)\\
$(id_x+d_y-d_z)/\sqrt{3} $&$\langle \mathbf{L} \rangle =\hbar (0,1,1)2/3\ $ & (cyclic)\\
$(d_x+\omega^2 d_y+\omega d_z)/\sqrt{3}\ $&$\langle \mathbf{L} \rangle =\hbar (1,1,1)/\sqrt{3}$& \\
$(\omega^2 d_y+\omega d_z)/\sqrt{2}$&$\langle \mathbf{L} \rangle =\hbar (1,0,0)\sqrt{3}/2\ $ &(cyclic) \\
\hline
\hline
\end{tabular}
\end{center}
  \caption{\label{lincom} The expectation values the angular momentum operator $\mathbf{L}$ for several complex linear combinations  of $t_{2g}$ orbitals. $\omega=e^{2\pi i/3}$}  
\end{table}

The quasi-particles $v^l_{\ssK\sigma}$ are expressed in terms of
pocket operators by $v^l_{\ssK\sigma}=\hat{U}^r_{lm}b^{m}_{\ssK
  \sigma}$, where the unitary matrix $\hat{U}^r_{lm}=U^r_{ln}\mF_{nm}$
diagonalizes $K^r$. From this follows that if $K^r$ is already
diagonal, no folding of the BZ occurs and translational symmetry is
not broken.  Otherwise, the BZ is folded and states of different
pockets recombine to form the new quasi-particles.

Now we consider finite expectation values of the spin-density
operators $\mathbf{S}^r$. Due to the absence of spin-orbit coupling,
our model has an $SU(2)$ rotational symmetry in spin space. Therefore
the discussion can be restricted to the order parameters $\langle
S^r_z\rangle=\langle\mathbf{e}_z\cdot\mathbf{S}^r\rangle$, given by
\begin{equation}
\label{smodes}
\langle S^r_z\rangle= \frac{2}{N}\sum_{\ssK \sigma}\lambda^r_l \sigma\langle v^{\dag l}_{\ssK\sigma}v^l_{\ssK\sigma}\rangle,
\end{equation}
where $\sigma$ takes the values $1$ and $-1$ corresponding to spin up
and down. If only one $\langle S^r_z\rangle\neq 0$, the effective
interaction Hamiltonian reduces to
\begin{equation}
\label{smeanmodes}
\frac{\Lambda^{\mathrm{s}}_r\langle S^r_z \rangle}{2} \sum_{\ssK \sigma}\lambda^r_l \sigma \; v^{\dag l}_{\ssK\sigma}v^l_{\ssK\sigma}.
\end{equation}
The mean-field Hamiltonian (\ref{smeanmodes}) is given by the same
quasi-particles and the same eigenvalues $\lambda^r_l$ as the
Hamiltonian in (\ref{meanmodes}). The only difference is that the sign
of the splitting of the quasi-particle bands depends on the spin.  In
the following, all ordering transitions with order parameters $\langle
n^r\rangle$ and $\langle S^r_z \rangle$ for $r=0,\dots,15$ are shortly
discussed.
\subsection*{r=0}
{\it charge:} $\langle n^0 \rangle$ is the total charge of the system,
which is fixed and non-zero, even in the paramagnetic phase.

{\it spin:} A finite $\langle S^0_z \rangle$ describes a Stoner
ferromagnetic instability.  The coupling constant
$\Lambda^{\mathrm{s}}_0$ given in TABLE~\ref{Gaal} is the most
negative coupling constant. In the unperturbed system without trigonal
distortion, the critical temperature of all continuous transitions
discussed here, only depends on the density of states and on the
coupling constant in the mean-field picture. In this case,
ferromagnetism is the leading instability for the unperturbed system.
In the real CoO$_2$-plane, this must not necessarily occur, but strong
ferromagnetic fluctuations will be present in any case.
 \subsection*{r=1--3}
\begin{figure}
 \begin{tabular*}{\linewidth}{|c@{\extracolsep{\fill}}c|}
 \hline
 \rule{0cm}{1.7cm}\includegraphics[height=0.17\linewidth]{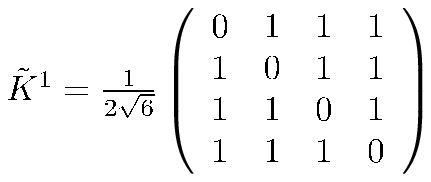}&
 \includegraphics[height=0.17\linewidth]{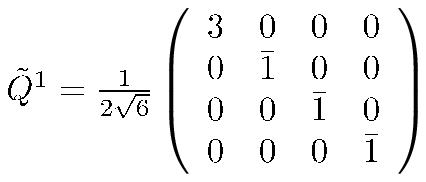}\\
 \includegraphics[width=0.42\linewidth]{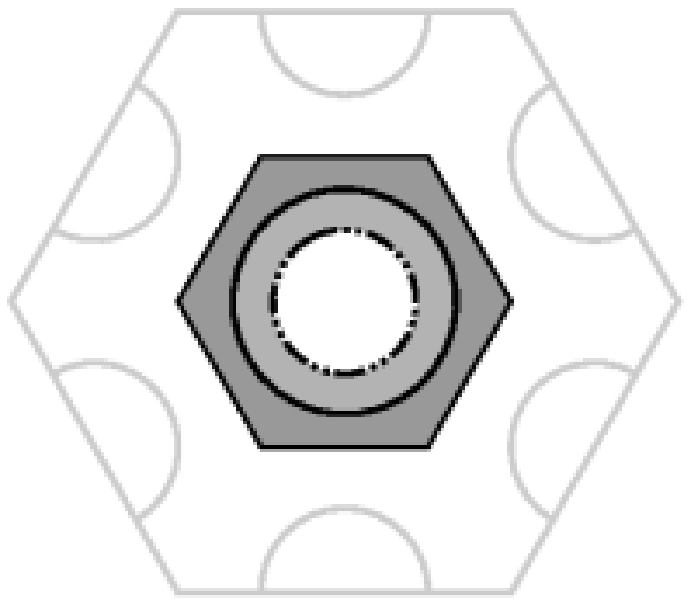}
&
 \includegraphics[height=0.38\linewidth]{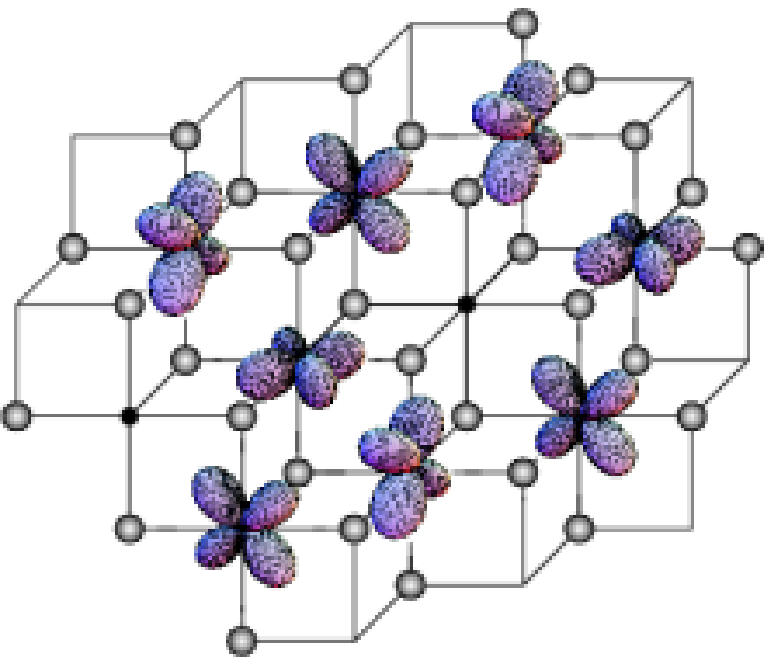}\\
 \hline
 \end{tabular*}
 \caption{\label{so} Charge ordering instability with finite expectation value $\langle \tilde{n}^1_{\scriptscriptstyle \mathbf{0}}\rangle $ leading to a charge enhancement or reduction on one Kagom\'e lattice. The folding of the BZ and the splitting of the pockets is shown (The double dotted line in the BZ indicates a triply degenerate pocket).
   On the right a quasi-particle state that is in this case just a
   Kagom\'e lattice state is drawn.}
 \end{figure}

 {\it charge:} A finite expectation value $\langle n^r \rangle$ for $
 r=1,2,3 $ corresponds to a difference in the charge density on the
 four Kagom\'e lattices, because the matrices $Q^{1-3}$ of
 TABLE~\ref{Q1-Q15} are diagonal and the quasi-particles $v^{\dag
   l}_{\ssK \sigma}$ are just the Kagom\'e states $a^{\dag l}_{\ssK
   \sigma}$. From the view point of Fermi surface pockets given by $
 K^{1-3} $ which are non-diagonal, this order yields a folding of the
 BZ, because the quasi-particles $v^{\dag l}_{\ssK \sigma}$ are linear
 combinations of states belonging to different hole pockets.  This
 means that the translational symmetry is broken. In the matrix
 $Q^{1-3}$ we find two positive and two negative diagonal elements.
 Consequently, a finite expectation value $\langle n^{1-3} \rangle$
 leads to a charge enhancement on two Kagom\'e lattices and to a
 charge reduction on the other two.  As specifying two Kagom\'e
 lattices specifies a direction on the triangular lattice, rotational
 symmetry is broken and crystal symmetry is reduced from hexagonal to
 orthorhombic.  The phases described by the matrices $Q^{1-3}$ have
 the same coupling constant $\Lambda^{\mathrm{c}}_1$ because they
 transform irreducibly into each other under crystal symmetries with
 the representation $\Gamma^a_5$. In order to examine which linear
 combinations of the three order parameters $\langle n^1 \rangle$,
 $\langle n^2 \rangle$ and $\langle n^3 \rangle$ could be stable below
 the critical temperature, we consider the Landau expansion of the
 free energy $\Delta F=F-F_0$
\begin{eqnarray}\label{landauI}
\Delta F&=&\frac{\alpha}{2}(\eta_1^2 +\eta_2^2 +\eta_3^2) + \beta\,\eta_1\eta_2\eta_3 +\frac{\gamma_1}{4}(\eta_1^2+\eta_2^2+\eta_3^2)^2\nonumber\\&&+\frac{\gamma_2}{4}(\eta_1^2\eta_2^2+\eta_2^2\eta_3^2+\eta_3^2\eta_1^2),
\end{eqnarray}
with $\eta_1=\langle n^1 \rangle$, $\eta_2=\langle n^2 \rangle$,
$\eta_3=\langle n^3 \rangle$.  For
$\gamma_1>\mathrm{max}\{0,-\gamma_2\}$, the free energy is globally
stable. For $\gamma_2<0$, Eq.~(\ref{landauI}) has a minimum of the
form $\eta_1=\eta_2=\eta_3$, if
$\beta^2-4\alpha(3\gamma_1+\gamma_2)>0$. This phase is described by
the symmetric combination $\tilde{Q}^1=(Q^1+Q^2+Q^3)/\sqrt{3}$ which
does not break the rotational symmetry.  In FIG.~\ref{so}, the folding
of BZ and the splitting of the bands (the dotted line is triply
degenerate) and the orbital pattern of the quasi-particles $v^{\dag
  l}_{\ssK\sigma}=a^{\dag l}_{\ssK\sigma}$ are shown. Note that
$\tilde{Q}^1$ has one positive and three negative diagonal elements.
The charge is enhanced or reduced on a single Kagom\'e lattice
depending on the sign of the coefficient $\beta$ in
Eq.~(\ref{landauI}). The third-order term in the free energy expansion
is allowed by symmetry, because there is no inversion-like symmetry
that would switch the of $(\eta_1,\eta_2,\eta_3) \to
(-\eta_1,-\eta_2,-\eta_3)$. Therefore the transition can be first
order.  On the other hand, for $\gamma_2>0$, there is a competition
between the terms proportional to $\gamma_2$ and $\beta$ in
Eq.~(\ref{landauI}). The minimum has not a simple form.  For
$|\beta|\ll \gamma_2$, however, the transition yields states
approximately described by the matrix $Q^1$, $ Q^2 $ or $ Q^3 $. In
any case this phase does break the rotational symmetry.

{\it spin:} The spin density mean-fields $\langle S^i_z \rangle$
$i=1,2,3$ transform under space group symmetries like $\Gamma_5^a$ and
time reversal symmetry gives $\langle S^i_z \rangle$ to $-\langle
S^i_z \rangle$.  Due to the latter the third order term in
Eq.~(\ref{landauI}) is forbidden, so that the transition is
continuous.  For $\gamma_2<0$, Eq.~(\ref{landauI}) has again a minimum
of the form $\eta_1=\eta_2=\eta_3$, whereas for $\gamma_2>0$ the
minimum is realized for $ \eta_1 \neq 0 $ and $\eta_2=\eta_3=0$ (and
permutations) , if $\alpha<0$.  The folding of the BZ, the
quasi-particles and the breaking of space-group symmetries is the same
as for the charge density operators $n^i$. However, the splitting of
the bands depends now on the spin and time reversal symmetry is
broken.

\begin{figure}
\begin{tabular}{cc}
 \rule{0cm}{1.7cm}\includegraphics[width=0.5\linewidth]{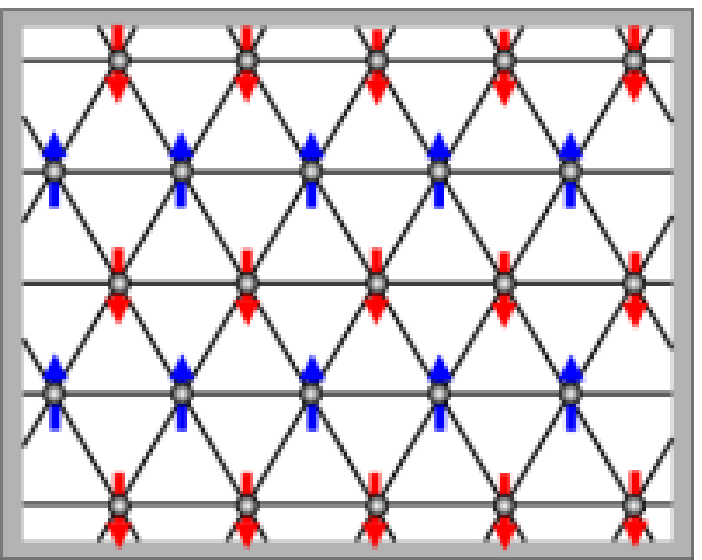}&
 \includegraphics[width=0.5\linewidth]{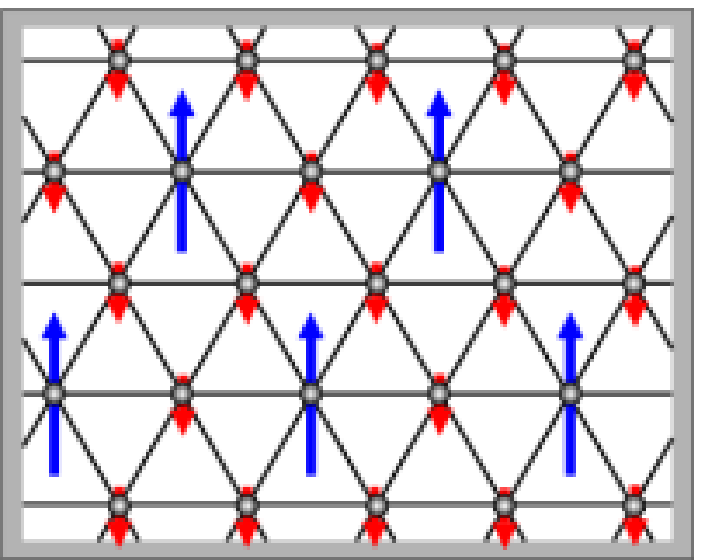}
\end{tabular}
 \caption{\label{sdw}The  spin density wave patterns corresponding to a finite expectation value $\langle S^1_z\rangle$ is shown on the left. This pattern is stabilized if $\gamma_2>0$ in Eq.~(\ref{landauI}). The pattern on the right corresponds to a finite order parameter $\langle S^1_z+S^2_z+S^3_z\rangle$, which is  stabilized for $\gamma_2<0$.
 }
 \end{figure}

These states are spin density waves, spatial modulations of the spin
density with a vanishing total magnetization. The two different types
of spin density modulations for $ \gamma_2 > 0 $ or $ \gamma_2 < 0 $
are shown in Fig.~\ref{sdw}.  For $\gamma_2>0$ rotational and
translational symmetry is broken yielding a collinear spin orientation
along one spatial direction and alternation perpendicular.  In
contrast $\gamma_2>0$ yields a rotationally symmetric spin density
wave with a doubled unit cell. This special type of spin density wave
gives a subset of lattice points, forming a triangular lattice, of
large spin density and another subset with opposite spin density of a
third in size, forming a Kagom\'e lattice.  Both states are metallic,
because no gaps are opened at the FS. This spin density wave is not a
result of Fermi surface nesting, but due to the complex orbital
structure.  The coupling constant for this transition,
$\Lambda^{\mathrm{s}}_1$ is the second strongest coupling in the model
Hamiltonian after the ferromagnetic coupling constant,
$\Lambda^{\mathrm{s}}_{0}$, as it is best seen in
Fig.~\ref{couplingplot}.
 \begin{figure}
\begin{tabular}{|c|}
\hline
 \begin{tabular*}{\linewidth}{l@{\extracolsep{\fill}}r}
 \rule{0cm}{0.5cm}\rule{1.65cm}{0cm}\includegraphics[height=0.05\linewidth]{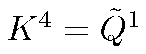}&
 \includegraphics[height=0.05\linewidth]{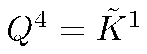}\rule{1cm}{0cm}\\
 \LARGE{a)}\includegraphics[height=0.35\linewidth]{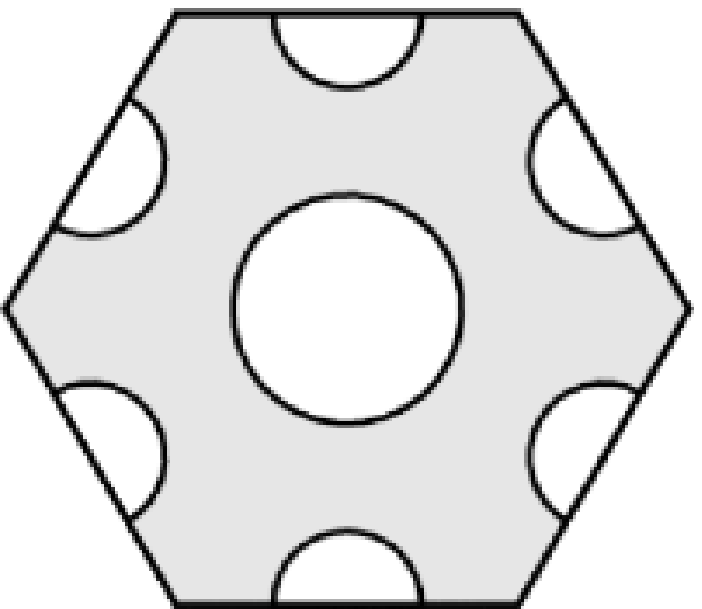}&
 \includegraphics[height=0.35\linewidth]{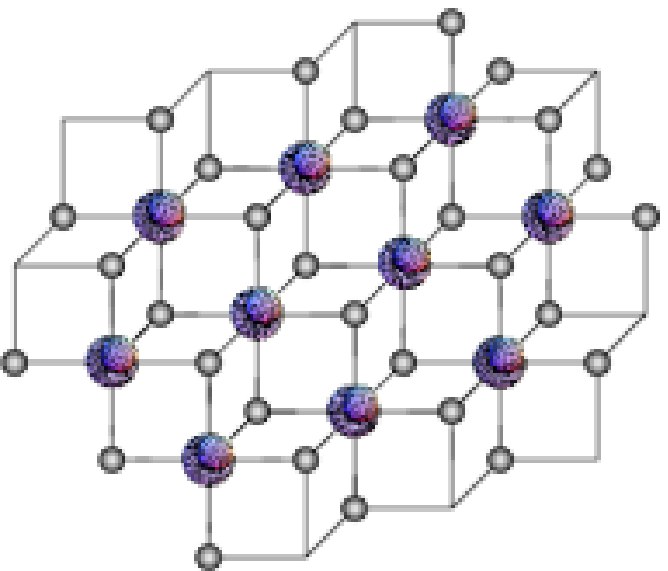}\\
\end{tabular*}\\
\hline\\
 \begin{tabular*}{\linewidth}{l@{\extracolsep{\fill}}r}
 \rule{0cm}{0.0cm}\rule{2.05cm}{0cm}\includegraphics[height=0.04\linewidth]{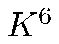}&
 \includegraphics[height=0.05\linewidth]{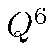}\rule{1.8cm}{0cm}\\
 \LARGE{b)}\includegraphics[height=0.35\linewidth]{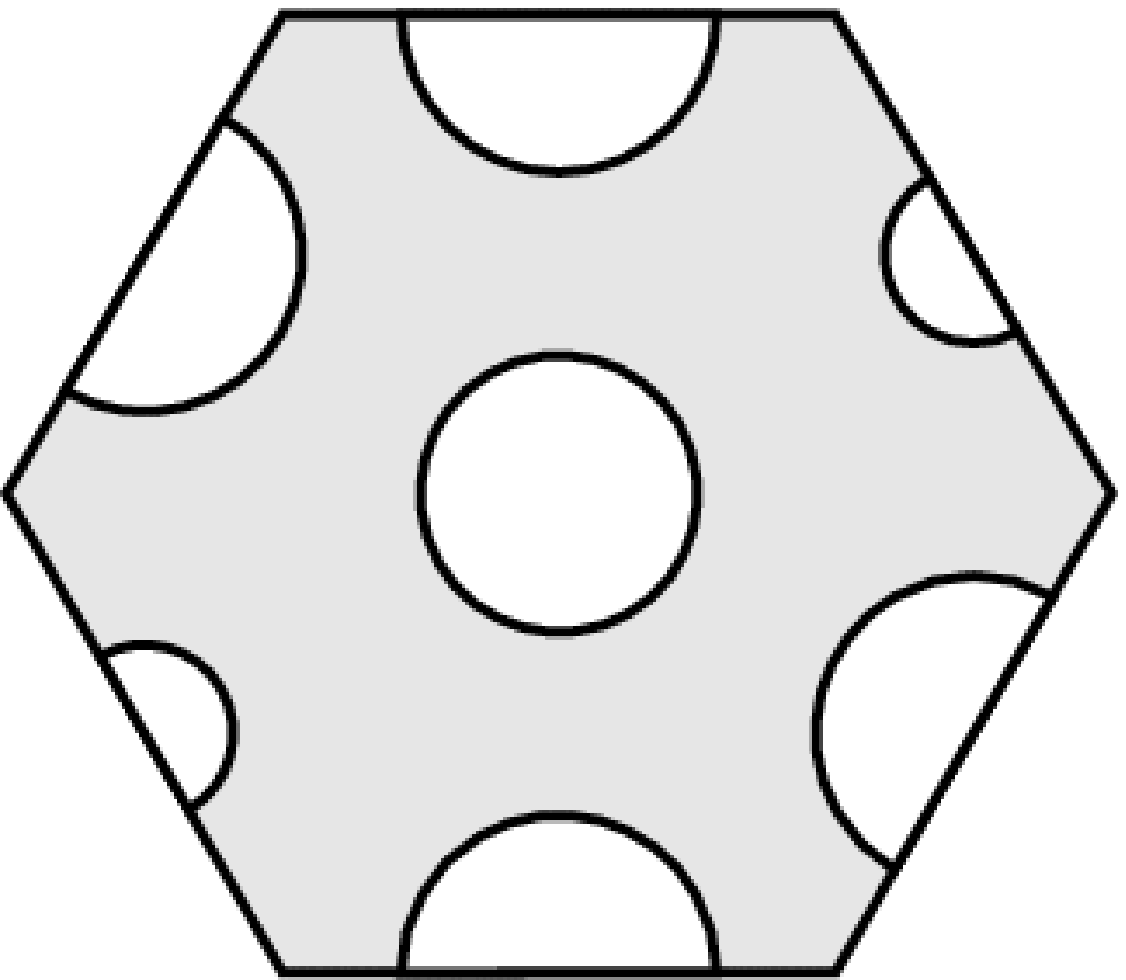}&
 \includegraphics[height=0.34\linewidth]{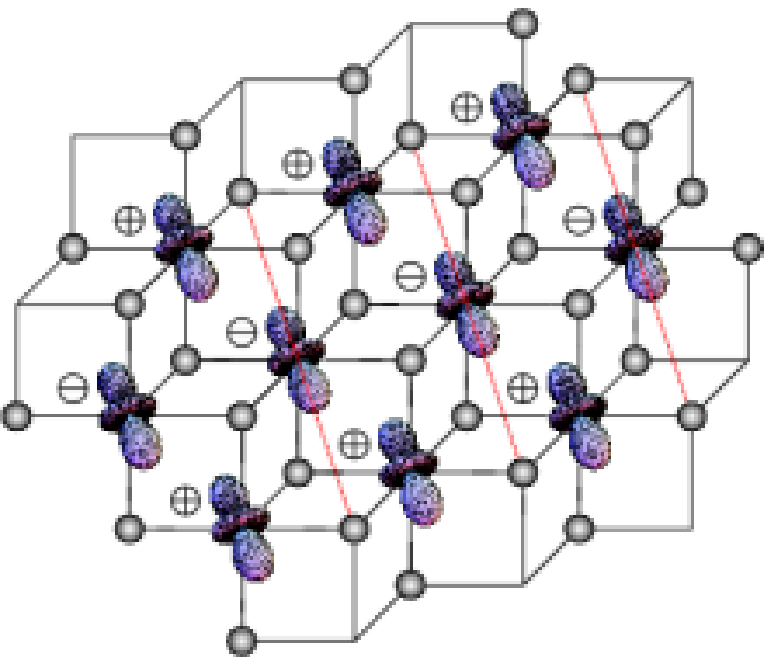}\\
\end{tabular*}\\
\hline\\ 
 \begin{tabular*}{\linewidth}{l@{\extracolsep{\fill}}r}
 \LARGE{c)}
\includegraphics[width=0.35\linewidth]{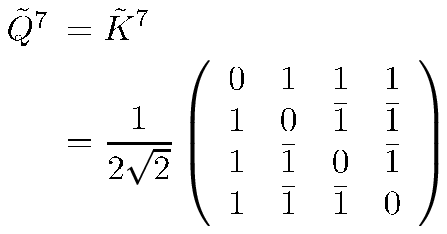}
&
 \includegraphics[height=0.4\linewidth]{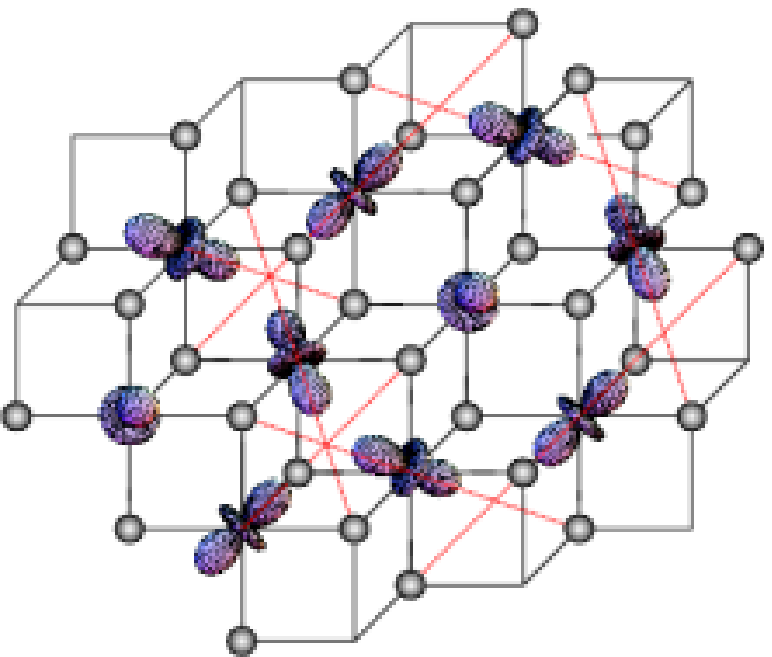}\\
\end{tabular*}\\
\hline
 \end{tabular}
 \caption{\label{bo}Ordering instabilities, described by real off-diagonal $SU(4)$ generators $Q^{4-9}$. $a)$ $Q^4$ breaks neither translational nor rotational symmetry. $b)$ shows the ordering corresponding to $Q^6$. The ordering shown in $c)$ is described in real space and in reciprocal space by the same matrix $\tilde{Q}^7=\tilde{K}^7$ and breaks translational symmetry. The corresponding BZ is shown in FIG.~\ref{so}.}
 \end{figure}
\subsection*{r=4}
{\it charge:} As discussed in section~\ref{breaksym}, a finite
expectation value of $n^4$ does not break any space group symmetry.
The matrix $K^4$ is diagonal with one positive and three negative
elements. This leads to a change of the band energy of the band at the
$\Gamma$ point relative to those at the M points (Fig.~\ref{bo}a)).
This results in an orbital order, a pattern as shown in
Fig.~\ref{bo}a), because the number of holes associated with the hole
pocket around the $ \Gamma $-point is different from that of the other
pockets. The net charge onsite vanishes, but the charge distribution
has the quadrupolar form, which results from
\begin{equation} \begin{array}{l}
\displaystyle \rho(r) \propto \frac{1}{4}[3 |\psi_{yz} + \psi_{zx} + \psi_{xy}|^2 - |\psi_{yz} - \psi_{zx} + \psi_{xy}|^2 \\ \displaystyle \qquad -
|\psi_{yz} + \psi_{zx} -\psi_{xy}|^2  -|\psi_{yz} - \psi_{zx} - \psi_{xy}|^2 ] \\
\displaystyle \quad = \psi^*_{yz} \psi_{zx} 
+ \psi^*_{zx} \psi_{xy} + \psi^*_{xy} \psi_{yz}  +c.c. \;.\
\end{array}
\label{charge-quadro-1}
\end{equation}
The corresponding tensor operator belongs to the representation $
\Gamma_1 $ of the subgroup $ D_3 $ of the cubic group with the
three-fold rotation axis parallel to [111], i.e. along the c-axis
perpendicular to layer. This quadrupolar field would be driven by the
symmetry reduction discussed above, through trigonal distortion and
direct $dd$-hopping among the $ t_{2g}$-orbitals.

{\it spin:} While the corresponding order parameter $\langle
S^4_z\rangle$ breaks time reversal symmetry, space group symmetry is
conserved.  This order is spatially uniform analogous to a ferromagnet
without, however, having a net magnetic moment. Because the magnetic
moment associated with the Fermi surface pocket at the $\Gamma$-point
is opposite and three times larger than the moment at the three
$M$-pockets.  While the net dipole moment vanishes on every site, this
configuration has a finite quadrupolar spin density corresponding to
the onsite spin density distribution of the same form as the charge
distribution in Eq.(\ref{charge-quadro-1}), which also belongs to $
\Gamma_1 $ representation of $ D_3 $.  It is also important to note
that no third order terms are allowed due to broken time reversal
symmetry, such that the transition to this order would be continuous.
\subsection*{r=5,6}
{\it charge:} The order parameters $\langle n^5\rangle=\eta_5$ and
$\langle n^6\rangle=\eta_6$ transform according to the irreducible
representation $\Gamma_3$ of the cubic point group. The Landau
expansion of the free energy is given by
\begin{equation}\label{landauII}
\Delta F=\frac{\alpha}{2}(\eta_5^2 +\eta_6^2) + \frac{\beta}{3}\eta_6(3\eta_5^2-\eta_6^2) +\frac{\gamma}{4}(\eta_5^2+\eta_6^2)^2,
\end{equation}
whose global stability requires $\gamma>0$. The third order term,
allowed here, induces a first order transition and simultaneously
introduce an anisotropy which is not present in the second- and
fourth-order terms. We can write $ (\eta_5,\eta_6) = \eta (\cos
\varphi , \sin \varphi ) $ and obtain
\begin{equation}
\Delta F = \frac{\alpha}{2} \eta^2 + \frac{\beta}{3} \eta^3 \sin 3 \varphi + \frac{\gamma}{4} \eta^4
\end{equation}
Depending on the sign of $ \beta $ the stable angles will be $ \varphi
= {\rm sign}(\beta) \pi /2 + 2 \pi n /3 $. This yields three
degenerate states of uniform orbital order whose charge distribution
has the quadrupolar form:
\begin{equation} \begin{array}{l} \displaystyle 
\rho(r) \propto e^{i\varphi} \{ (\psi^*_{zx} \psi_{xy} + \psi^*_{xy} \psi_{zx}) + \omega (\psi^*_{yz} \psi_{zx}
+ \psi^*_{zx} \psi_{yz} ) \\ 
\displaystyle + \omega^2 (\psi^*_{xy} \psi_{yz} + \psi^*_{yz} \psi_{xy}) \} + c.c.
\label{charge-quadro-2}
\end{array}
\end{equation}
with a tensor operator belonging to $ \Gamma_3 $ of $ D_3 $.  Each
state is connected with the choice of one $ M $-pocket which has a
different filling compared to the other two (Fig. ~\ref{bo}b).  The
main axis of each state points locally along one of the three cubic
body-diagonals, $[\bar{1},1,1]$, $[1,\bar{1},1]$, $[1,1,\bar{1}]$, and
the sign of the local orbital wave function is staggered along the
corresponding direction on the triangular lattice, $[\bar{2},1,1]$,
$[1,\bar{2},1]$, $[1,1,\bar{2}]$. In this way the rotational symmetry
is broken but the translational symmetry is conserved. The matrices
$Q^5$ and $Q^6$ commute with $Q^4$ such that the external symmetry
reduction has only a small effect on this type of order.

{\it spin:} The spin densities $\langle S^5_z\rangle$ and $\langle
S^6_z\rangle$ also belong to the two-dimensional representation $
\Gamma_3 $ of the cubic point group. Here time reversal symmetry
ensures that the Landau expansion only allows even orders of the order
parameter $ (\eta_5, \eta_6 ) = \eta (\cos \varphi, \sin \varphi) $
(i.e. $\beta = 0 $ in Eq.(\ref{landauII}).  The continuous degeneracy
in $ \varphi $ is only lifted by the sixth order term, given by
\begin{equation}\label{landauIII}
\frac{\delta_1}{6}(\eta_1^2 +\eta_2^2)^3 + \frac{\delta_2}{6}\eta_2^2(3\eta_1^2-\eta_2^2)^2 = 
\frac{\delta_1}{6} \eta^6 + \frac{\delta_2}{6} \eta^2 \sin^2 3 \varphi
\end{equation}
Stability requires $\delta_1>\mathrm{max}\{0,-\delta_2\}$. The
anisotropy is lifted by the $ \delta_2 $-term which give rise to two
possible sets of three-fold degenerate states.  Depending on the sign
of $ \delta_2 $ we have a minimum of the free energy for $ \varphi =
(1 - {\rm sign} \delta_2) \pi/4 + \pi n $. The corresponding spin
densities have no net dipole on every site, but again a quadrupolar
form of the same symmetry as for the charge, given by
Eq.(\ref{charge-quadro-2}).

\subsection*{r=7-9}
{\it charge:} The order parameters $\langle n^i \rangle$ for $i=7,8,9$
transform irreducibly under space group symmetries with the
representation $\Gamma_5^b$. The expansion~(\ref{landauI}) of the free
energy holds also for these order parameters. The third order term
makes the transition first order and favors the symmetric rotationally
invariant combination of the order parameters, described by the matrix
$\tilde{Q}^7=(Q^7+Q^8+Q^9)/\sqrt{3}=\tilde{K}^7$ shown in
FIG.~\ref{bo}c). The folding of the BZ and the splitting of the bands
is the same as in Fig.~\ref{so}.  The orbital pattern of the
non-degenerate quasi-particle band is also shown in Fig.~\ref{bo}c).
It consists of atomic $\varphi^0$ orbitals pointing along all four
cubic space diagonals.  Translational but not rotational symmetry is
broken.

{\it spin:} The discussion for the spin density operators is analogous
to the discussion in the section $\bf{r=1-3}$.
 \begin{figure}
 \begin{tabular*}{\linewidth}{|l@{\extracolsep{\fill}}r|}
 \hline
 \rule{0cm}{1.7cm}\includegraphics[height=0.18\linewidth]{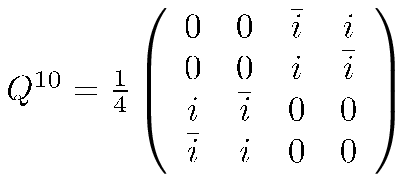}&
 \includegraphics[height=0.17\linewidth]{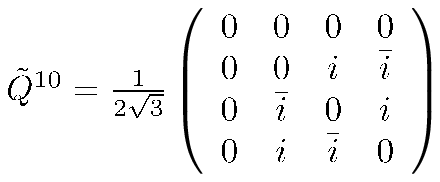}\\
 \LARGE{a)}\includegraphics[height=0.35\linewidth]{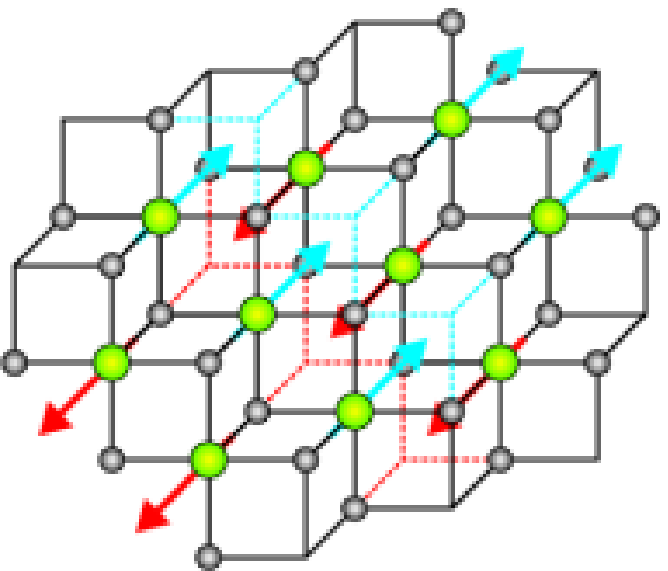}&
 \includegraphics[height=0.35\linewidth]{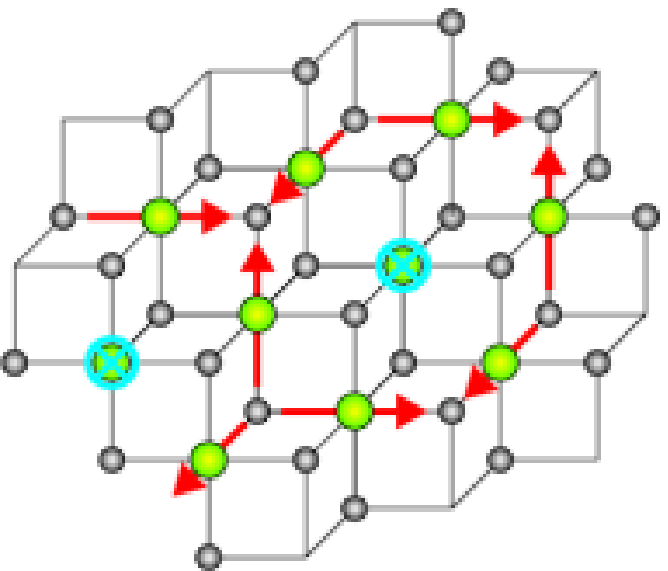}\\
 \hline
 \rule{0cm}{1.7cm}\rule{0.0cm}{0cm}\includegraphics[height=0.18\linewidth]{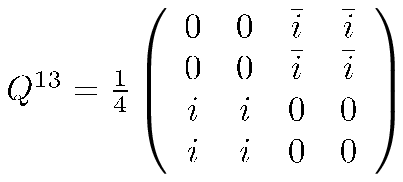}&
 \rule{0cm}{1.7cm}\includegraphics[height=0.18\linewidth]{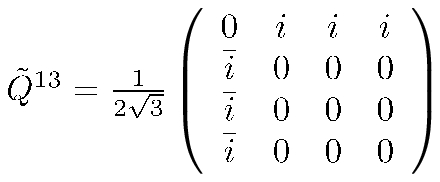}\rule{0.0cm}{0cm}\\
 \LARGE{b)}\includegraphics[height=0.35\linewidth]{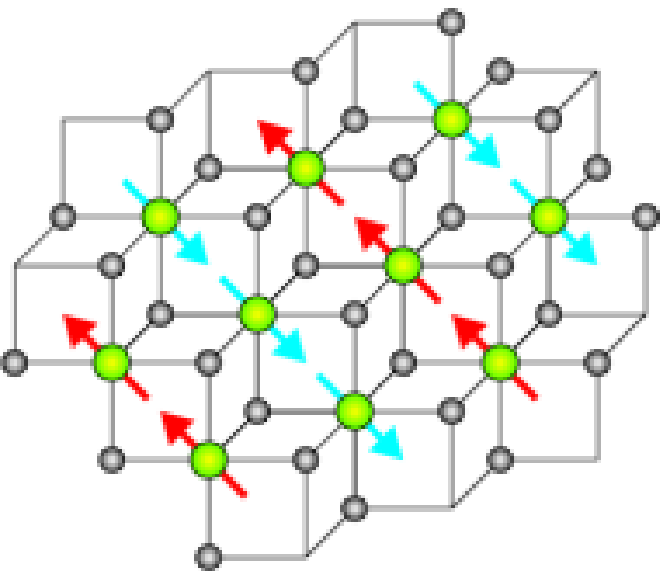}&
 \includegraphics[height=0.35\linewidth]{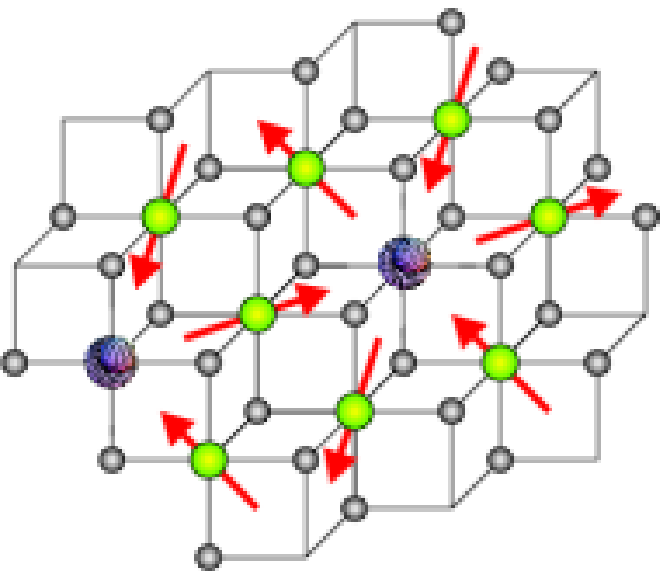}\\
 \hline
 \LARGE{c)}\rule{0cm}{3.2cm}\rule{0.2cm}{0cm}\includegraphics[height=0.35\linewidth]{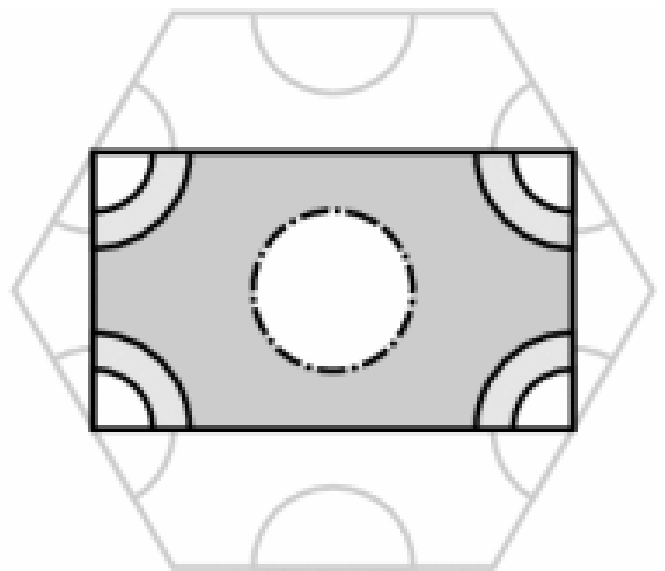}&
 \includegraphics[height=0.35\linewidth]{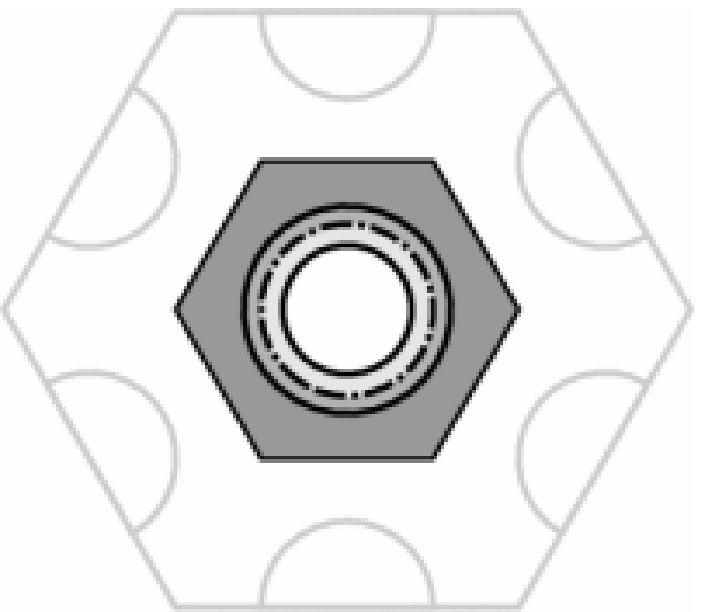}\\
 \hline
 \end{tabular*}
 \caption{\label{co}Transitions to time reversal symmetry breaking states, where the expectation value of the orbital angular momentum $\langle \vec{L}\rangle$ on the Co-sites is finite. $a)$ States where the angular momentum does not lie in the plane. $\tilde{K}^{10}=\tilde{Q}^{10}$ $b)$ States with angular momentum in the plane. $\tilde{K}^{13}=-\tilde{Q}^{13}$. $c)$ The folding of the BZ and the hybridization of the bands for $a)$ are shown. Dotted lines indicate double degenerate bands.}
 \end{figure}
\subsection*{r=10-12}
{\it charge:} The order parameters $\langle n^i \rangle$ for
$i=10,11,12$ transform irreducibly under space group symmetries with
the representation $\Gamma_4$. For the $\Gamma_4$ representation of
$T_d$, there is no third order invariant. All other terms in
Eq.~(\ref{landauI}) are however also invariants for $\Gamma_4$. The
absence of the third order term leads to continuous transition.  The
stabilized state for $\alpha<0$ depends on the sign of $\gamma_2$ in
Eq.~(\ref{landauI}).

For $\gamma_2>0$ a nontrivial minimum with $\langle n^{11}
\rangle=\langle n^{12} \rangle=0$ exists, which is described by the
hermitian, imaginary matrix $Q^{10}$.  If $\lambda$ is an eigenvalue
of $Q^{10}$, then $-\lambda$ is also an eigenvalue of $Q^{10}$ and the
corresponding quasi-particles are connected by time reversal symmetry.
Therefore the non-vanishing eigenvalues of $Q^{10}$ belong to
quasi-particle states, which are not invariant under time reversal
symmetry. They are given by complex linear combinations of $t_{2g}$
orbitals. For complex linear combinations of $t_{2g}$ orbitals, the
expectation value of the orbital angular momentum operator
$\langle\mathbf{L}\rangle$ does not vanish in general, as can be seen
from TABLE~\ref{lincom}. In FIG.~\ref{co}a), the pattern of the
angular momentum expectation values $\langle\mathbf{L}\rangle$ for a
quasi-particle of $Q^{10}$ is shown. It is invariant under
translations along $\ba_1$ and staggered under translations along
$\ba_2$ and $\ba_3$. The expectation values are parallel to $[011]$.
The folding of the BZ and the splitting of the bands is shown in
Fig.~\ref{co}c). Rotational, translational and time reversal symmetry
is broken and the state has the magnetic point-group
$\underline{2}/\underline{m}$.

For $\gamma_2<0$ the symmetric combination
$\tilde{Q}^{10}=(Q^{10}+Q^{11}+Q^{12})/\sqrt{3}=\tilde{K}^{10}$ is
stabilized.  The angular momentum pattern for a quasi-particle with
non-vanishing eigenvalue is shown in FIG.~\ref{co}a). Depending on the
site, the expectation value points along $[100]$, $[010]$, $[001]$ or
$[\bar{1}\bar{1}\bar{1}]$ and the magnitudes are such, that the
pattern is rotationally invariant and the expectation value of the
total angular momentum perpendicular to the plane vanishes. The
folding of the BZ and the splitting of the pockets is shown in
FIG.~\ref{co}c). This state has the magnetic point group
$\underline{\overline{3}}\underline{m}$.  Note, that these states can
also be considered as a kind of staggered flux states. The matrices
$Q^{10-12}$ commute with $Q^4$ and therefore the transitions are only
little affected by a trigonal distortion.

{\it spin:} The spin density order parameters $\langle S^i_z\rangle$
for $i=10,11,12$ also transform under space group symmetries like
$\Gamma_4$ and except for the spin dependent quasi-particle energy,
the discussion is the same as for the charge density operators.  Note,
however, that these spin density operators do not change sign under
time reversal symmetry, because both the orbital angular momentum and
the spin is reversed.  This, however does not lead to a third order
term in the Landau expansion, as there is no third order invariant for
the $\Gamma_4$ representation anyway.
\subsection*{r=13-15}
{\it charge:} The order parameters $\langle n^i \rangle$ for
$i=13,14,15$ transform irreducibly under space group symmetries with
the representation $\Gamma_5^c$. The matrices $Q^{13-15}$ are also
imaginary and time reversal symmetry changes the sign of the order
parameters. The Landau expansion of the free energy is given as above
by Eq.~(\ref{landauI}) with $\beta=0$.

For $\gamma_2>0$ and $\alpha<0$ a minimum of the free energy is given
by the order parameter $\langle n^{13}\rangle$. The angular momentum
pattern of the quasi-particles is shown in FIG.~\ref{co}b). The
expectation values lie in the CoO$_2$-plane and are parallel to the
$\ba_1$ direction. Their sign is staggered along the $\ba_2$ and
$\ba_3$ direction. The quasi-particles consist of states belonging to
the $\Gamma$ and the M pocket. The folding of the BZ is given in
FIG.~\ref{co}c), but with the single dotted line in the center being a
doubly degenerate M pocket. Rotational, translational and time
reversal symmetry is broken.

For $\gamma_2<0$ the symmetric combination
$\tilde{Q}^{13}=(Q^{13}+Q^{14}+Q^{15})/\sqrt{3}=-\tilde{K}^{13}$ is
stabilized. The pattern of the quasi-particles corresponding to
$\tilde{Q}^{13}$ is shown in FIG.~\ref{co}b). It consists of
non-magnetic sites with a $\varphi_0$ orbital perpendicular to the
plane and of sites with angular momentum expectation values along
$\ba_i$. Rotational symmetry is not broken in this case. The folding
of the BZ and the splitting of the bands is shown in FIG.~\ref{co}c).

All angular momentum expectation values for these two states lie in
the CoO$_2$-plane. Therefore, it is not possible to interpret these states
as staggered flux states.

{\it spin:} The spin density order parameters $\langle S^i_z \rangle$
for $i=13,14,15$ are invariant under time reversal symmetry.
Therefore, the third order term in Eq.~(\ref{landauI}) is allowed and
the transition is a first order transition.
 \section{Possible Instabilities\label{possinst}} 
 
 \subsection{coupling constants}
 As can be seen from TABLE~\ref{Gaal}, the coupling constants for the
 SDW transitions $\Lambda^{\mathrm{s}}$ are rather negative whereas
 the charge coupling constants $\Lambda^{\mathrm{c}}$ tend to be
 positive. This is not surprising as only local repulsive interaction
 is considered here, that tends to spread out the charge as much as
 possible.
 
 The coupling constants $\Lambda^{\mathrm{c/s}}_r$ with $r=0,\dots,3$
 depend on the intra-orbital Coulomb repulsion $U$. As $U$ is the
 largest Coulomb integral, the absolute value of these coupling
 constants is biggest.  The remaining coupling constants
 $\Lambda^{\mathrm{c/s}}_r$ do not depend on $U$. For $J'=0$ they are
 also independent of $r$. For finite $J'$ the degeneracy between the
 real (4-9) and imaginary (10-15) SU(4) generators is lifted.

\begin{figure}[htbp]
   \includegraphics[width=\linewidth]{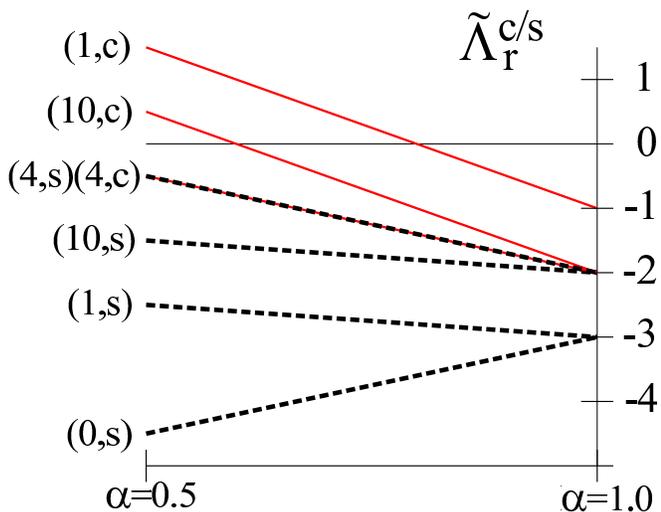}
   \caption{The dimensionless coupling constants
     $\tilde{\Lambda}^{\mathrm{c/s}}_{r}=9\Lambda^{\mathrm{c/s}}/(2U)$
     as functions of $\alpha=U'/U$. The relations $U=U'+2\Jh$ and
     $J'=\Jh$ are assumed to hold. The solid (dashed) lines denote the
     charge (spin) coupling constants.
   \label{couplingplot}}
 \end{figure}

 In order to compare the different coupling constants better, the
 relations $U=U' +2\Jh$ and $\Jh=J'$, that hold in spherically
 symmetric system, can be assumed to hold approximately. The ratio
 $\alpha=U'/U$ is positive and usually larger than 1/2 and smaller
 than 1. These assumptions allow to order the dimensionless coupling
 constants $\tilde{\Lambda}^{\mathrm{c/s}}=
 9\Lambda^{\mathrm{c/s}}/(2U)$ according to their strength. In
 FIG.~\ref{couplingplot} the dimensionless coupling constants
 $\tilde{\Lambda}^{\mathrm{c/s}}_r$ are shown as functions of
 $\alpha$.
 The most negative coupling constant is the ferromagnetic one with
 $\tilde{\Lambda}^{\mathrm{s}}_0=-6+3\alpha$. For $\alpha$ close to 1,
 the coupling constant for spin density order
 $\tilde{\Lambda}^{\mathrm{s}}_1=-(2+\alpha)$ is comparable.  Smaller
 but still clearly negative are also the coupling constants for the
 spin density angular momentum states
 $\tilde{\Lambda}^{\mathrm{s}}_{10}=-(1+\alpha)$.  The coupling
 constants
 $\tilde{\Lambda}^{\mathrm{c}}_{4}=\tilde{\Lambda}^{\mathrm{s}}_{4}=1-3\alpha$
 are also negative.  Finally, the coupling constant for time reversal
 symmetry breaking angular momentum states
 $\tilde{\Lambda}^{\mathrm{c}}_{10}=3-5\alpha$ and for the charge
 density order $\tilde{\Lambda}^{\mathrm{c}}_1=4-5\alpha$ are rather
 positive, but can in principle also be negative if $\alpha$ is close
 enough to one.  In fact it is quite remarkable that for $\alpha>0.8$
 all coupling constants constants (except $\Lambda^{\mathrm{c}}_0$)are
 negative.  For $\alpha=1$ additional degeneracies among the coupling
 constants appear, as can be seen in FIG.~\ref{couplingplot}.  This
 indicates the existence of a higher symmetry at this point. In fact,
 the local Coulomb interaction $H^{\mathrm{C}}_{\mathbf{r}}$ of
 Eq.~\ref{inter} depends only on the total charge
 $n_{\mathbf{r}}=\sum_{m\sigma}n_{\mathbf{r}m\sigma}$ on the site
 $\mathbf{r}$ and is given by $Un_{\mathbf{r}}(n_{\mathbf{r}}-1)/2$
 for $\alpha=1$.

\subsection{effect of the trigonal distortion}
In the mean-field description, an instability occurs if the
Stoner-type criterion is satisfied.  At zero temperature in the system
with full symmetry, this criterion reads in our notation as
\begin{eqnarray}
  \label{stonercr}
 \frac{ -\Lambda^{\mathrm{c/s}}_r}{4}D(E_{\mathrm{F}})=1,
\end{eqnarray}
where $D(E_{\mathrm{F}})$ is the density of states per spin and per
hole pocket. For rather small pockets $D(E_{\mathrm{F}})$ is given by
$\sqrt{3}/(2\pi t)\approx 0.28/t$ in our tight binding model,
increases however with decreasing $E_{\mathrm{F}}$ (cf.\ 
Fig.~\ref{zones}).  From Eq.~(\ref{stonercr}) we can estimate that the
critical $U$ must be larger than $10t$ for having a ferromagnetic
instability. With the introduction of the trigonal distortion, as it
was discussed in section~\ref{breaksym}, the Stoner criteria of
Eq.~(\ref{stonercr}) are modified.

For the order parameters described by the matrices $K^0$,
$K^4$,$K^{5-6}$ and $K^{10-12}$, that commute with the trigonal
distortion $K^{4}$, the change of the Stoner criterion is only due to
the changing of the density of states at the M and the $\Gamma$
pockets by the trigonal distortion, and the Stoner criterion is only
slightly modified as long as all four pockets exist.

On the other hand, the instabilities towards states, where the order
parameters with the matrices $K^{13-15}$ are finite, would be strongly
affected by the trigonal distortion, as the pocket states that
hybridize in such a transition are no longer degenerate.

Finally, as mentioned above, the order parameters described by the
matrices $K^{1-3}$ and $K^{7-9}$ transform with the same
representation and are mixed by the trigonal distortion.  For strong
distortions the mixing tends to odd-even combinations and only the odd
combinations, $K^{1}-K^{7}$, $K^{2}-K^{8}$, $K^{3}-K^{9}$ commute with
the symmetry breaking field, $K^4$, and connects the still degenerate
states of the M pockets.

If the trigonal distortion is so strong, that the pockets states at
the M points lie below the FS, only a spontaneous ferromagnetic
instability can still occur according to the Stoner criterion.  First
order transitions, however, are still possible.
     
The ferromagnetism is the leading instability in the symmetric model
and is least affected by the trigonal distortion.  Therefore, in real
Na$_x$CoO$_2$ systems where a rather strong trigonal distortion is
unavoidably present, ferromagnetism would be most robust and is in
fact the only type of all the described, exotic symmetry breaking
states, that would have a chance to occur spontaneously.

However, even if the coupling constants of the more exotic states are
not negative enough, to produce a spontaneous instability, their
corresponding susceptibilities can be large enough to give rise to an
important response of the electrons in CoO$_2$-plane to external
perturbations.  In the next section, we describe how te Na-ions can be
viewed as an external field for the charge degrees of freedom.

 \section{Na-superstructures\label{superstructure}}
 
 In Na$_{x}$CoO$_2$ the Na-ions separate the CoO$_2$ planes.  There
 are two different Na-positions which are both in prismatic
 coordination with the nearest O-ions.  The Na2 position is also in
 prismatic coordination with the nearest Co-ions, while the Na1
 position lies along the $c$-axes between two Co-ions below and above.
 This leads to significant Na--Co repulsion, suggesting that the Na1
 position is higher in energy.  In fact, the Na2 position is the
 preferred site for Na$_{0.75}$CoO$_2$, where the ratio of occupied
 Na1-sites to occupied Na2-sites is about 1:2.\cite{huang2}
 Deintercalation of Na does however not lead to a further depletion of
 the Na1-sites. On the contrary, the occupancy ratio goes to 1 for $x$
 going to 0.5.  Further there is a clear experimental evidence, that
 at $x=0.5$ the Na-ions form a commensurable orthorhombic
 superstructure already at room temperature.\cite{huang} For several
 other values of $x$ also superstructure formation has been reported,
 but $x=0.5$ shows the strongest signals and has the simplest
 superstructure.\cite{zandbergen,shi} In addition for $x=0.5$ samples
 a sharp increase of the resistivity at 50K respectively at 30K was
 reported.\cite{foo,chen,chen2}
 
 This experimental situation is rather surprising, naively one expects
 commensurability effects to be strongest at $x=1/3$ or at $x=2/3$ on
 a triangular lattice but not at $x=1/2$. Therefore, it was concluded
 that structural and electronic degrees of freedom are coupled in a
 subtle manner in Na$_x$CoO$_2$.\cite{huang2}
 
\begin{figure}
  \includegraphics[width=\linewidth]{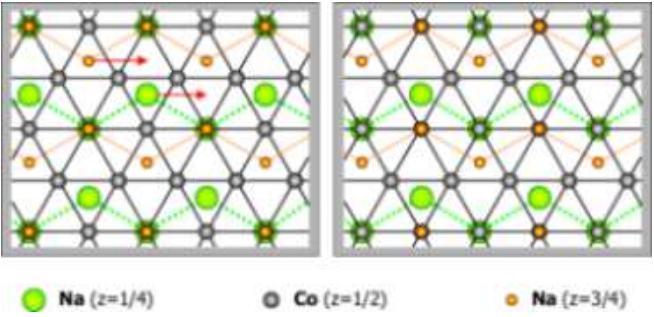}
 \caption{\label{Napatterns}
   Two different Na superstructures in Na$_{1/2}$CoO$_2$. The left one
   does not break rotational symmetry and would drive a charge
   ordering as shown in FIG.~\ref{bo}c).  The right one is in fact
   realized in Na$_{1/2}$CoO$_2$, it is obtained from the right one by
   shifting the Na-chains along the arrows. This shift is due to the
   Coulomb repulsion of the Na-ions.  }
 \end{figure}

 In this section we show how  the different ordering patterns can couple
to  the observed Na superstructure 
at $x=1/2$.  Before going into the details, we
 note that due to our starting point of inter-penetrating Kagom\'e
 lattices, commensurability effects will be strongest for samples
 where the Na-ions can form a simple periodic superstructures that
 double or quadruple the area of the unitcell, since specifying a
 single Kagom\'e lattice also quadruples the unitcell.  For $x=1/2$
 such simple superstructures exist as shown in FIG.~\ref{Napatterns}.
 A sodium superstructure couples to the charge but not to the spin
 degrees of freedom in the CoO$_2$ layer.
\begin{figure}
  \includegraphics[width=\linewidth]{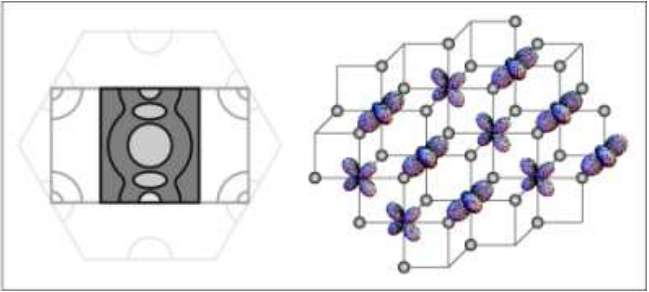}
 \caption{\label{Naunitc}  The charge ordering pattern corresponding to the matrix $K^1-K^7$, consists of alternating rows of $d_x$ and $d_y-d_z$ orbitals. On the left hand side the original BZ, the orthorhombic BZ due to the charge ordering and the experimentally observed reduced orthorhombic BZ (dark) are shown. 
 }
 \end{figure}
 In our model, there are 15 collective charge degrees of freedom.
 From FIG.~\ref{couplingplot} can be seen that
 $\Lambda_r^{\mathrm{c}}$ is most negative for $r=4,\dots,9$. Hence,
 these modes are the ``softest'' charge modes generating the strongest
 response to a Na-pattern. As shown in FIG.~\ref{bo}, the charge order
 corresponding to $r=4,5,6$ does not enlarge the unitcell and does
 therefore not optimally couple to the Na-patterns that can be formed
 with $x=0.5$.  However the orbital pattern shown in FIG.~\ref{bo}c)
 has lobes of electron density pointing towards selected Na1 and Na2
 positions. For $x=1/2$ it is possible to occupy all these and only
 these positions. This leads to the left Na-superstructure of
 FIG.~\ref{Napatterns}. In other words, this Na-superstructure couples
 in a optimal way to this rotationally symmetric charge pattern.
 Further, the Landau expansion shows that the rotationally symmetric
 combination is favored by the third-order term.  Therefore, it is
 clear that the electronic degrees of freedom would favor this
 Na-superstructure.  This pattern however does not maximize the
 Na-ions distances. It is apparent that the average distances between
 the sodium ions can be increased, if every second of the
 one-dimensional sodium chains is shifted by one lattice constant as
 shown in FIG.~\ref{Napatterns}. In this way an orthorhombic
 Na-superstructure is obtained, which is the one observed in
 experiments.  This orthorhombic pattern does not drive the
 rotationally symmetric charge pattern shown in FIG.~\ref{bo}c), which
 is described by the matrix $\tilde{K}^7=K^7+K^8+K^9$, it might
 however drive the orthorhombic charge pattern described by the matrix
 $K^7$ or rather the orthorhombic charge pattern described by
 $K^1-K^7$, as in the presence of trigonal distortion the $K^1$ and
 $K^7$ mix and the odd combination will have the most negative
 coupling constant. This charge pattern is shown in
 FIG.~\ref{Naunitc}.  It consists of lines of $d_x$-orbitals
 alternating with lines of the linear combination $d_y-d_z$-orbitals.
 Note, that this charge pattern corresponds to the mixed
 $K^1-K^7$-matrix, the charge is not uniformly distributed on the
 Co-atoms.  In this charge pattern, the Na1-sites above the
 $(d_y-d_z)$ Co-sites will be lower in energy than the Na1-sites above
 the $d_x$ Co-sites and similarly the Na2-positions are separated into
 nonequivalent rows.
 
 In reciprocal space, such a charge ordering leads to a folding of the
 BZ such that the two M-pockets hybridize. The ordering of the
 Na-ions along the chains leads to a further folding of the BZ and to
 a hybridization of the bands, as it is shown in FIG.~\ref{Naunitc}.
 The schematic FS in FIG.~\ref{Naunitc} is drawn to illustrate the
 hybridization occurring due to the translational symmetry breaking.
 Li et al.\ performed density-functional calculations in order to
 determine the band-structure of Na$_{0.5}$CoO$_2$ in the presence of
 the orthorhombic superstructure from first-principles.\cite{li} Quite
 generally one can assume that this superstructure, which specifies a
 direction on the triangular lattice, can lead to quasi-onedimensional
 bands in the reduced BZ.  For such one-dimensional bands, nesting
 features are likely to occur and would lead to a SDW-like
 instability, as it was observed at 53 K by Huang et
 al..\cite{huang,uemura} Such a transition could open a gap at least
 on parts of the FS and in this way lead to the drastic increase of
 the resistivity observed at 53 K.\cite{foo} At higher temperature,
 the resistivity is comparable in magnitude to the metallic samples
 and increases only slightly with lowering temperature.  This weakly
 insulating behavior could be another effect of Na-ion ordering. Since
 the rotational symmetry is broken, domains can be formed. The
 existence of domain walls would be an obstacle for transport where
 thermally activated tunneling processes play a role.  It would be
 interesting to test this idea by removing the domains and see whether
 metallic temperature dependence of the resistivity would result. A
 bias on the domains can be given by in-plane uniaxial distortion.
 
 To finish this section, we will discuss a further mechanism, that
 could lead to a non-magnetic low-temperature instability in
 Na$_{0.5}$CoO$_2$. In section \ref{OP} we saw that the third order
 term in the Landau expansion, Eq.~(\ref{landauI}), favors always a
 rotationally symmetric charge ordering where all three order
 parameters $\eta_1$, $\eta_2$ and $\eta_3$ have the same magnitude.
 But as argued above, the Na-ion repulsion leads nevertheless to an
 orthorhombic charge ordering, where only one order parameter $\eta_1$
 is finite.  From Eq.~(\ref{landauI}), we obtain a Landau expansion
 for the remaining two order parameters $\eta_2$ and $\eta_3$
 containing only second and forth order terms.  The second order term
 is given by
\begin{equation}
  \label{lastlandau}
 \frac{\tilde{\alpha}}{2}(\eta_2^2+\eta_3^2)+\tilde{\beta}\eta_2\eta_3,
\end{equation}
where
\begin{equation}
\label{tildea}
\tilde{\alpha}=\alpha+\left(\gamma_1+\frac{\gamma_2}{2}\right)\eta_1^2,\qquad \tilde{\beta}=\beta\eta_1.
\end{equation} 
The condition for a second order phase transition, that leads to
finite values of $\eta_2$ and $\eta_3$ is
$\tilde{\alpha}<|\tilde{\beta}|$.  As we have $\alpha>0$ and linear
growth of $|\tilde{\beta}|$ and quadratic growth of
$\tilde{\alpha}-\alpha$ with $\eta_1$, the condition is fulfilled
neither for large nor for small values of $\eta_1$.  But for
intermediate values of $\eta_1$ it can be fulfilled.  This tendency
back towards the original hexagonal symmetry in this or a similar form
could be responsible for the appearance of additional Bragg peaks at
the intermediate temperature of 80-100K in
Na$_{0.5}$CoO$_2$.\cite{huang} Note however, that it was speculated
that these Bragg peaks only exist over a narrow range of temperature.
\section{Conclusions\label{discussion}}
In this paper the properties of a high-symmetry multi-orbital
model for the CoO$_2$-layer in combination with local
Coulomb interaction are discussed. The tight-binding model is a zeroth
order approximation to the kinetic energy, as it only includes the
most relevant hopping processes using Co-O $\pi$-hybridization.
  Nevertheless it produces the hole
pocket with predominantly $a_{1g}$ character around the $\Gamma$
point, in agreement with both LDA calculations and ARPES experiments.
Furthermore, the three further pockets around the M points, although
not seen in ARPES experiments, suggest that additional degrees of
freedom that can not be captured in a single-band picture could be
relevant.  The existence of identical hole, pockets in the BZ does
however not produce pronounced nesting features.

The local Coulomb repulsion of the $t_{2g}$-orbitals can be taken into
account by an effective interaction of fermions with four different
flavors, associated with the four hole pockets or the four
inter-penetrating Kagom\'e lattices.  This effective interaction has a
large discrete symmetry group, that allows to classify the spin- and
charge-density operators, and to the determine for every mode the
corresponding coupling constant.

It turns out that with an effective trigonal distortion, that splits
the degeneracy between the $\Gamma$ and the M points, general
corrections to the quadratic part of the Hamiltonian, such as trigonal
distortion or additional hopping terms, can be taken into account,
provided they are small.  This effective trigonal distortion reduces
the symmetry of the Hamiltonian down to the space-group symmetries of
the CoO$_2$-plane, by breaking the gauge symmetries of the effective
interaction.
   
Most coupling constants are negative for reasonable assumptions on the
Coulomb integrals $U$, $U'$, $J'$ and $\Jh$, but the ferromagnetic
coupling constant is most negative and constitutes the dominant
correlation.  The charge and spin density wave instabilities without
trigonal distortion are easily described in a mean-field picture. In
reciprocal space the degenerate bands split, and if bands belonging to
different pockets hybridize, the BZ is folded.  In real space
different types of orderings are possible. The occupancy of the
different $t_{2g}$ orbitals on different sites can be nonuniform,
resulting in a charge ordering with nonuniform charge distribution on
the Co-sites. Further, certain real or complex linear combinations of
$t_{2g}$ orbitals can be preferably occupied on certain sites. In this
case, the charge is uniformly distributed on the sites, but depending
on the linear combinations of the orbitals, certain space group
symmetries are broken.  The complex linear combinations of $t_{2g}$
orbitals have in general a non-vanishing expectation value of the
orbital angular momentum.

The tendency to these rather exotic states turns out to be smaller
than the ferromagnetic tendency, and this dominance of the
ferromagnetic state is even more enhanced by the trigonal distortion.
This is in good agreement with experiments, where ferromagnetic
in-plane fluctuations have been observed by neutron scattering
measurements in Na$_{0.75}$CoO$_2$.\cite{boothroyd,helme}  There are
also several reports of a phase transition in Na$_{0.75}$CoO$_2$ at
22K to a static magnetic order, which is probably ferromagnetic
in-plane but antiferromagnetic along the
$c$-axis.\cite{motohashi,sugiyama1,sugiyama2}

In Na$_{0.5}$CoO$_2$, a periodic Na-superstructure couples directly to
a charge pattern in our model and crystallizes already at room
temperature, whereas simple $\sqrt{3}\times\sqrt{3}$-superstructures,
that would correspond $x=1/3$ or $x=2/3$ do not couple.

For general values of $x$ the disordered Na-ions provide a random
potential that couples to the charge degrees of freedom. Due to the
incommensurability, this does not lead to long range order, but the
short range correlations will also be influenced by the charge degrees
of freedom in the CoO$_2$ layers.  This interaction between the
Na-correlations and the charge degrees of freedom could be the origin
of the charge ordering phenomena at room temperature and the
observation of inequivalent Co-sites in NMR
experiments.\cite{ray,gavilano}


The overall agreement of our model with the experimental situation is
good.  Ferromagnetic fluctuations are dominant in our model and in
experiments.  Furthermore, our model is based on a metallic state and
allows for charge ordering and spin density ordering transitions
without changing the metallic character of the state.  Finally, the
clear Na superstructures, that were found at $x=0.5$, can be
understood quite naturally in this model.

On the other hand there are still many open questions for the
cobaltates.  Mainly the origin and the symmetry of the superconducting
state of Na$_x$CoO$_2\cdot$yH$_2$O is still under debate.
Unfortunately the Na content $x=0.35$ is beyond the validity  of the approximations
made in the derivation of  
our model.  But also the samples with
$x\geq0.5$ have still many intriguing properties like the strongly
anisotropic magnetic susceptibility, which shows the unusual
Curie-Weiss temperature dependence.
A possible description of the anisotropy of the magnetic susceptibility could be achieved  by introducing 
a spin-orbit term into the kinetic energy.

We hope that our model
will be useful for further understanding of the rich experimental situation of
the cobaltates.
\section{acknowledgments}
We thank E.\ Bascones, B.\ Batlogg, M.\ Br\"uhwiler, J.\ Gavilano, H.R.\ Ott, T.M.\ Rice, and K.\ Wakabayashi for fruitful discussions.
This work is financially supported by a grant of the Swiss National Fund and the NCCR program MaNEP of the Swiss National Fund.

 \appendix
 \section{}\label{tbappend}
 
 The equivalence of the two definitions for the "pocket-operators"
 made in Eq.~(\ref{kpocketop}) and in Eq.~(\ref{kkbgoop}) follows from
 \begin{eqnarray}\label{2defb}
 b^{\dag j}_{\ssK m}&=&\frac{1}{2}\sum_le^{i\ssB_j\cdot\ba_l}a^{\dag l}_{\ssK m}\nn
 &=&\frac{1}{2}\sum_l e^{i\ssB_j\cdot \ba_l}\frac{2}{\sqrt{N}}\sum_{\ssR}e^{i\ssK\cdot(\ssR+\ba_l+\ba_m)}a^{\dag l}_{\ssR m}\nn
 &=&e^{-i\ssB_j\cdot\ba_m}\frac{1}{\sqrt{N}}\sum_{l\ssR}e^{i(\ssK+\ssB_j)\cdot(\ssR+\ba_l+\ba_m)}c^{\dag}_{\ssR+\ba_l+\ba_m m}\nn
 &=&e^{-i\ssB_j\cdot\ba_m}c^{\dag}_{\ssK +\ssB_j m}
 \end{eqnarray}
 The diagonal form of the tight-binding Hamiltonian in
 Eq.~(\ref{abtbham}) follows directly from the relation
 \begin{equation}
   \label{diag}
   \epsilon^{mm'}_{\ssK +\ssB_j}=e^{-i\ssB_j\cdot(\ba_m-\ba_{m'})}\epsilon_{\ssK}^{mm'}.
 \end{equation}

 \section{}\label{deriv}
 In this appendix, we provide some details concerning the derivation
 of the effective Hamiltonian in Eq.~(\ref{charsp}).  It is convenient
 to treat each term in Eq.~(\ref{inter}) separately. Let us start with
 the Hund's coupling.
 \begin{eqnarray}\label{firststeps}
 &&\frac{\Jh}{2}\sum_{{\mathrm r}}\sum_{m\neq m'}c^{\dag}_{\br m\s}c^{\dag}_{\br m' \s'}c_{\br m\s'}c_{\br m'\s} \\
 &=&\frac{\Jh}{2N}\sum_{\mathbf{kqk'q'}}^r\sum_{m\neq m'}c^{\dag}_{\mathbf{k} m\s}c^{\dag}_{\mathbf{k'} m' \s'}c_{\mathbf{q} m\s'}c_{\mathbf{q'} m'\s}\label{ema2}\\
 &=&\frac{\Jh}{2N}\sum_{\ssK\ssK'\ssQ}\sum_{ijkl}^r \sum_{m\neq m'}e^{i{\scriptscriptstyle (\ssB_i-\ssB_k)}\cdot\ba_m}e^{i{\scriptscriptstyle (\ssB_l-\ssB_j)}\cdot\ba_{m'}}\times\nn
 &&\qquad \times b^{\dag i}_{{\scriptscriptstyle\ssK} m\s}b^{\dag l}_{{\scriptscriptstyle -\ssK+\ssQ}\, m' \s'}b^k_{{\scriptscriptstyle -\ssK'+\ssQ}\, m\s'}b^j_{{\scriptscriptstyle\ssK'} m'\s}\label{ema3}
 \end{eqnarray}
 The sum over the momenta in (\ref{ema2}) is restricted such that
 $\mathbf {k+k'-q-q'}$ equals a reciprocal lattice vector.
 (\ref{ema3}) follows from (\ref{ema2}) by using the definition of the
 pocket operators in Eq.~(\ref{kkbgoop}).  The sum over the pocket
 indices is again restricted such that $\sB_i+\sB_j+\sB_k+\sB_l$
 equals a reciprocal lattice vector, whereas the sum over the momenta
 in the reduced BZ is simplified to an unrestricted sum over three
 momenta.  Note that this simplification is valid for small pockets,
 because all the processes at the Fermi energy are kept.  (Small
 pockets means here $4\sK_{\mathrm{F}}<|\sB_1|$, this corresponds to a
 doping with $x>0.55$) The next step is to go from orbital operators
 to the band operators. Restricting ourselves to the top band and
 taking into account Eq.~(\ref{symab}) we can simply substitute
 $b^{\dag j}_{\ssK m \s}\rightarrow \frac{1}{\sqrt{3}} b^{\dag
   j}_{\ssK\s}$. Now we can sum over the orbital indices in
 Eq.~(\ref{ema3}) and taking into account that the sum over the pocket
 indices is restricted we obtain the sum
 \begin{equation}
   \label{summmp}
   \sum_{m\neq m'}e^{i(\ssB_i-\ssB_k)\cdot(\ba_m-\ba_{m'})}=2(4\delta_{ik}-1)
 \end{equation}
 and for the Hund's coupling term
 \begin{equation}\label{subsu}
 \frac{\Jh}{9N}\sum_{\ssK\ssK'\ssQ}\sum_{ijkl}^r b^{\dag i}_{{\scriptscriptstyle\ssK} \s}b^{\dag l}_{{\scriptscriptstyle -\ssK+\ssQ}\,  \s'}b^k_{{\scriptscriptstyle -\ssK'+\ssQ}\, \s'}b^j_{{\scriptscriptstyle\ssK'} \s}(4\delta_{ik}-1)
 \end{equation}
 The restriction of the sum can be dropped, if we replace
 $(4\delta_{ik}-1)$ with
 $(2\delta_{ijkl}-\epsilon^2_{ijkl}-\delta_{il}\delta_{jk}-\delta_{ij}\delta_{kl}+3\delta_{ik}\delta_{jl})$.
 The terms proportional to $\Jh$ in the interaction of
 Eq.~(\ref{charsp}) are now obtained by dividing Eq.~(\ref{subsu})
 into two equal parts, rewrite one directly in terms of density
 density operators, and rewrite the other in terms of density-density
 and spin-density spin-density operators using the $SU(2)$ relation
 $2\delta_{\alpha\delta}\delta_{\beta\gamma}=\delta_{\alpha\gamma}\delta_{\beta\delta}+\boldsymbol{\sigma}_{\alpha\gamma}\cdot\boldsymbol{\sigma}_{\beta\delta}$.
 Terms which renormalize the chemical potential are dropped. All the
 other terms in Eq.~(\ref{inter}) are treated in the same way.

 \section{}\label{QandG}
 The symmetry group $G$ of $H_{\mathrm{eff}}$ is a finite subgroup of
 $U(4)$ that is generated by t the permutation matrices $\mP\in\mS_4$
 and the diagonal orthogonal matrices $\mD\in (Z_2)^4$.  $G$ is a
 semi-direct product of $S_4$ and the normal subgroup $(Z_2)^4$, this
 allows us to find the irreducible representations of $G$,
 cf.\cite{serre} The elements can be written in a unique way as
 $(\mP,\mD)$ with $\mP\in\mS_4$ and $\mD\in (Z_2)^4$. The product of
 two elements $(\mP,\mD)\circ (\mP',\mD')$ is given by
 $(\mP\circ\mP',\mD'')$. From this follows that if $(\mP,\mD)$ is
 conjugate to $(\mP',\mD')$, $\mP$ is conjugate to $\mP'$, and the
 class of $(\mP,\mD)\in G$ can be labelled by the class of
 $\mP\in\mS^4$. The elements of $S_4$ can be classified by writing
 them as disjunct cyclic permutations. We label the five classes as
 follows: $e$=$1$, $f$=$(ab)$, $g$=$(ab)(cd)$, $h$=$(abc)$,
 $i$=$(abcd)$.
 In total there are twenty 20 classes in $G$. The character table is
 shown in TABLE~\ref{chartab}. The character corresponding to the
 natural representation of $G$ by orthogonal $4\times 4$ matrices is
 $\chi_{11}$. The representation on the 16 dimensional space $V$
 spanned by $Q^{0-15}$, that was defined in section ~\ref{SU4order},
 acts irreducibly on the subspaces $V^0$, $V^{1-3}$, $V^{4-9}$ and
 $V^{10-15}$ with the characters $\chi_{0}$, $\chi_{7}$, $\chi_{15}$
 and $\chi_{16}$, respectively.
 
 With help of Schur's Lemma, it is now easy to show that the
 interaction $H_{\mathrm{eff}}$ in the Basis $Q^{0-15}$ is diagonal,
 i.e.
 \begin{equation}
   \label{Lamdef}
  Q^r_{ji}A^{\mathrm{c/s}}_{ijkl}Q^{r'}_{kl}=\delta_{rr'}\Lambda^{\mathrm{c/s}}_r.
 \end{equation}
 and that the coupling constant $\Lambda^{\mathrm{c/s}}_r$ depend only
 on the irreducible subspace.
 
 As discussed in section~\ref{breaksym}, the subgroup $(Z_2)^4$
 describes gauge-symmetries, that are broken in the real system
 whereas the subgroup $\mS_4$ describes the space-group symmetries.
 The subgroup $\mS_4$ consists of the classes $e_1$, $f_1$, $g_1$,
 $h_1$ and $i_1$. The irreducible representations of $G$ are in
 general reducible for the subgroup $\mS_4$. For example we have
 $\chi_7=\Gamma_5$, $\chi_{15}=\Gamma_1\oplus\Gamma_3\oplus\Gamma_5$
 and $\chi_{16}=\Gamma_4\oplus\Gamma_5$.

 \begin{table*}
 \begin{center}
 $\begin{array}[c]{l|ccccc|cccccc|ccc|cccc|cc|c|}
  & e_1& e_2& e_3& e_4& e_5& f_1& f_2& f_3& f_4&  f_5& f_6& g_1& g_2& g_3& h_1& h_2& h_3& h_4& i_1& i_2&\textrm{reduction}\\
  \#& 1& 4& 6& 4& 1& 12& 12& 24& 24& 12& 12& 12& 24& 12& 32& 32& 32& 32& 48& 48&\textrm{to}\quad \mS_4\\
 \hline
  \chi_{1}&      1&1&1&1&1&1&1&1&1&1&1&1&1&1&1&1&1&1&1&1& \Gamma_1\\
  \chi_{2}&      1&1&1&1&1&\bar{1}&\bar{1}&\bar{1}&\bar{1}&\bar{1}&\bar{1}&1&1&1&1&1&1&1&\bar{1}&\bar{1}&\Gamma_2 \\
  \chi_{3}&      1&\bar{1}&1&\bar{1}&1&1&\bar{1}&\bar{1}&1&1&\bar{1}&1&\bar{1}&1&1&\bar{1}&\bar{1}&1&1&\bar{1}&\Gamma_1 \\
  \chi_{4}&      1&\bar{1}&1&\bar{1}&1&\bar{1}&1&1&\bar{1}&\bar{1}&1&1&\bar{1}&1&1&\bar{1}&\bar{1}&1&\bar{1}&1 &\Gamma_2\\
  \chi_{5}&      2&2&2&2&2&0&0&0&0&0&0&2&2&2&\bar{1}&\bar{1}&\bar{1}&\bar{1}&0&0 &\Gamma_3\\
  \chi_{6}&      2&\bar{2}&2&\bar{2}&2&0&0&0&0&0&0&2&\bar{2}&2&\bar{1}&1&1&\bar{1}&0&0 &\Gamma_3\\
  \chi_{7}&      3&3&3&3&3&1&1&1&1&1&1&\bar{1}&\bar{1}&\bar{1}&0&0&0&0&\bar{1}&\bar{1}&\Gamma_5 \\
  \chi_{8}&      3&3&3&3&3&\bar{1}&\bar{1}&\bar{1}&\bar{1}&\bar{1}&\bar{1}&\bar{1}&\bar{1}&\bar{1}&0&0&0&0&1&1 &\Gamma_4\\
  \chi_{9}&      3&\bar{3}&3&\bar{3}&3&1&\bar{1}&\bar{1}&1&1&\bar{1}&\bar{1}&1&\bar{1}&0&0&0&0&\bar{1}&1 &\Gamma_5\\
  \chi_{10}&     3&\bar{3}&3&\bar{3}&3&\bar{1}&1&1&\bar{1}&\bar{1}&1&\bar{1}&1&\bar{1}&0&0&0&0&1&\bar{1} &\Gamma_4\\
  \chi_{11}&     4&2&0&\bar{2}&\bar{4}&2&2&0&0&\bar{2}&\bar{2}&0&0&0&1&\bar{1}&1&\bar{1}&0&0 & \Gamma_1\oplus\Gamma_5\\
  \chi_{12}&     4&2&0&\bar{2}&\bar{4}&\bar{2}&\bar{2}&0&0&2&2&0&0&0&1&\bar{1}&1&\bar{1}&0&0 &\Gamma_2\oplus\Gamma_4\\
  \chi_{13}&     4&\bar{2}&0&2&\bar{4}&2&\bar{2}&0&0&\bar{2}&2&0&0&0&1&1&\bar{1}&\bar{1}&0&0 &\Gamma_1\oplus\Gamma_5\\
  \chi_{14}&     4&\bar{2}&0&2&\bar{4}&\bar{2}&2&0&0&2&\bar{2}&0&0&0&1&1&\bar{1}&\bar{1}&0&0 &\Gamma_2\oplus\Gamma_4\\
  \chi_{15}&     6&0&\bar{2}&0&6&2&0&0&\bar{2}&2&0&2&0&\bar{2}&0&0&0&0&0&0 & \Gamma_1\oplus\Gamma_3\oplus\Gamma_5\\
  \chi_{16}&     6&0&\bar{2}&0&6&0&2&\bar{2}&0&0&2&\bar{2}&0&2&0&0&0&0&0&0&\Gamma_4\oplus\Gamma_5 \\
  \chi_{17}&     6&0&\bar{2}&0&6&0&\bar{2}&2&0&0&\bar{2}&\bar{2}&0&2&0&0&0&0&0&0 &\Gamma_4\oplus\Gamma_5\\
  \chi_{18}&     6&0&\bar{2}&0&6&\bar{2}&0&0&2&\bar{2}&0&2&0&\bar{2} &0&0&0&0&0&0 &\Gamma_1\oplus\Gamma_3\oplus\Gamma_5\\
  \chi_{19}&     8&4&0&\bar{4}&\bar{8}&0&0&0&0&0&0&0&0&0&\bar{1}&1&\bar{1}&1&0&0 &\Gamma_3\oplus\Gamma_4\oplus\Gamma_5\\
  \chi_{20}&     8&\bar{4}&0&4&\bar{8}&0&0&0&0&0&0&0&0&0&\bar{1}&\bar{1}&1&1&0&0&\Gamma_3\oplus\Gamma_4\oplus\Gamma_5
      \\ \hline
 \end{array}$
 \caption{\label{chartab} The character table for the symmetry group $G$ of the effective Hamiltonian $H_{\mathrm{eff}}$. The first line labels the classes and gives the number of elements in each class. The letters of the classes indicate classes of the subgroup $\mS_4$: $e$=$1$, $f$=$(ab)$, $g$=$(ab)(cd)$, $h$=$(abc)$, $i$=$(abcd)$. The characters appearing in our effective Hamiltonian are $\chi_1$ for $Q^0$, $\chi_7$ for $Q^{1-3}$, $\chi_{15}$ for $Q^{4-9}$ (real matrices) and $\chi_{16}$ for $Q^{10-15}$ (imaginary matrices). $\chi_{11}$ is the natural representation of $G$ defined in section~\ref{SU4order}. The last column gives the reduction of the representations into irreducible representations of the subgroup $\mS_4$, that consists of the classes $e_1$, $f_1$, $g_1$, $h_1$ and $i_1$.}
 \end{center}
 \end{table*}


\begin{thebibliography}{99}
\bibitem{tanakaT}
T.\ Tanaka et al., Jpn.\ J.\ Appl.\ Phys.\ \textbf{33},L581 (1994).
%
\bibitem{terasaki}
I.\ Terasaki et al., Phys.\ Rev.\ B \textbf{56},  R12685 (1997).
%
\bibitem{valla}
T.\ Valla et al., Nature {\bf 417}, 627 (2002).
%
\bibitem{wang1}
Q.\-H.\ Wang et al., Nature \textbf{423}, 425 (2003). 
%
\bibitem{takada} 
K.\ Takada et al., Nature {\bf 422}, 53 (2003).
%
\bibitem{schaak}
R.\-E.\ Schaak et al., Nature {\bf 424}, 527 (2003).
%
\bibitem{foo}
M.\-L.\ Foo et al., Phys.\ Rev.\ Lett.\ \textbf{92}, 247001(2004).
%
\bibitem{huang} 
Q.\ Huang et al., J.\ Phys.:\  Condens.\ Matter \textbf{16}, 5803 (2004).
%
\bibitem{zandbergen}
H.\-W.\ Zandbergen et al., Phys.\ Rev.\ B \textbf{70}, 024101 (2004).
%
\bibitem{shi}
Y.\-G.\ Shi et al., cond-mat/0401052.
%
\bibitem{chen}
X.Z.\ Chen et al., cond-mat/0412299. 
%
\bibitem{huang2}
Q.\ Huang et al., Phys.\ Rev.\ B \textbf{70}, 184110 (2004).
%
\bibitem{wangNL}
N.L.\ Wang et al., Phys.\ Rev.\ Lett.\ \textbf{93}, 237007 (2004).
%
\bibitem{ray}
R.\ Ray et al., Phys.\ Rev.\ B \textbf{59} 9454 (1999). 
%
\bibitem{gavilano}
J.\-L.\ Gavilano et al., Phys.\ Rev.\ B \textbf{69}, 100404(R) (2004).
%
\bibitem{carretta}
P.\ Carretta et al., Phys.\ Rev.\ B \textbf{70}, 024409 (2004). 
%
\bibitem{markus}
M.\ Br\"uhwiler et al., cond-mat/0309311.
%
\bibitem{bernhard}
C.\ Bernhard et. al., Phys.\ Rev.\ Lett.\ \textbf{93}, 167003 (2004).
%
\bibitem{lupi}
S.\ Lupi et al.,  Phys.\ Rev.\ B \textbf{69}, 180506R (2004).
%
\bibitem{balicas}
L.\ Balicas et al., cond-mat/0410400.
%
\bibitem{chen2}
X.H.\ Chen et al., cond-mat/0501181.
%
\bibitem{lupi2}
S.\ Lupi et al., cond-mat/0501746.
%
\bibitem{boothroyd}
A.\-T.\ Boothroyd et al., Phys.\ Rev.\ Lett.\ \textbf{92}, 197201
%
\bibitem{helme}
L.M.\ Helme et al., cond-mat/0410457.
%
\bibitem{ihara}
Y.\ Ihara et al., cond-mat/0407195.
%
\bibitem{motohashi}
T. Motohashi et al., Phys.\ Rev.\ B \textbf{67}, 064406 (2003).
%
\bibitem{sales}
B.\-C.\ Sales et al., Phys.\ Rev.\ B \textbf{70}, 174419 (2004).
%
\bibitem{sugiyama1}
J.\ Sugiyama et al., Phys.\ Rev.\ Lett.\ \textbf{92}, 017602 (2004).
%
\bibitem{sugiyama2}
J.\ Sugiyama et al., Phys.\ Rev.\ B \textbf{67}, 214420 (2003).
%
\bibitem{chou}
F.\-C. Chou et al., cond-mat/0404061.
%
\bibitem{caimi}
G.\ Caimi et al., Eur.\ Phys.\ J.\ B \textbf{40}, 231 (2004).
%
\bibitem{luo}
J.\-L.\ Luo et al.,  Phys.\  Rev.\  Lett.\ \textbf{93}, 187203 (2004).
%
\bibitem{uemura}
Y.J.\ Uemura et al., cond-mat/0403031 .
%
\bibitem{mendels}
P.\ Mendels et al., cond-mat/0501203.
%
\bibitem{miyoshi}
K.\ Miyoshi et al.,  Phys.\ Rev.\ B \textbf{69}, 132412 (2004).
%
\bibitem{baskaran1}
G.\ Baskaran, Phys. Rev. Lett. \textbf{91}, 097003 (2003).
%
\bibitem{kumar}
B.\ Kumar and B.\-S.\ Shastry, Phys.\ Rev.\ \textbf{B} 68, 104\,508 (2003).
%
\bibitem{tanaka}
A.\ Tanaka and X.\ Hu, Phys.\ Rev.\ Lett.\ \textbf{91}, 257006 (2003).
%
\bibitem{ogata}
M.\ Ogata, J.\ Phys.\ Soc.\ Jpn.\ \textbf{72}, 1839 (2003).
%
\bibitem{wang2}
Q.\-H.\ Wang et al., Phys. Rev. B \textbf{69}, 092504 (2004).
%
\bibitem{carsten}
C.\ Honerkamp , Phys.\ Rev.\ B \textbf{68}, 104510 (2003).
%
\bibitem{ferraz}
A.\ Ferraz et al., cond-mat/0412235.
%
\bibitem{maekawa}
W.\ Koshibae and S.\ Maekawa, Phys.\ Rev.\ Lett.\ \textbf{91}, 257003 (2003).
%
\bibitem{yanase}
Y.\ Yanase et al., cond-mat/0407563.
%
\bibitem{singh1}
D.\-J.\ Singh, Phys.\ Rev.\ B \textbf{61}, 13397 (2000).
%
\bibitem{singh2}
D.\-J.\ Singh, Phys.\ Rev.\ B \textbf{68}, 020503 (2003).
%
\bibitem{singh3}
M.\-D.\ Johannes et al., cond-mat/0403135.
%
\bibitem{kunes}
K.-W.\ Lee et al., Phys.\ Rev.\ B \textbf{70}, 045104 (2004).
%
\bibitem{li}
Z.\ Li et al., cond-mat/0403727 .
%
\bibitem{zhang}
P.\ Zhang et al., Phys.\ Rev.\ B \textbf{70},085108 (2004).
%
\bibitem{zhang2}
P.\ Zhang et al., cond-mat/0502072.
%
\bibitem{johannes}
M.D.\ Johannes et al., cond-mat/0408696.
%
\bibitem{hasan}
M.\-Z.\ Hasan et al., Phys.\ Rev.\ Lett.\ \textbf{92}, 246402 (2004).
%
\bibitem{yang}
H.-B.\ Yang et al., Phys.\ Rev.\ Lett.\ \textbf{92}, 246403 (2004).
%
\bibitem{yang2}
H.-B.\ Yang et al.,cond-mat/0501403.
\bibitem{wu}%
W.B.\ Wu et al., cond-mat/0408467.
%
\bibitem{serre}
 J.\-P. Serre, {\it Linear Representations of Finite Groups}, (Springer-Verlag, New York, 1977).
%
\end{thebibliography}
 \end{document}